\documentclass[]{aa}

\usepackage{txfonts}
\usepackage{natbib}
\bibpunct{(}{)}{;}{a}{}{,}
\usepackage{graphicx}
\usepackage {epstopdf}
\usepackage{caption}

\begin{document}

\title{A\,0535+26 in the April 2010 outburst: probing the accretion regime at
work}
\author{Daniela M\"uller \inst{\ref{inst1}}
\and Dmitry Klochkov \inst{\ref{inst1}}
\and Isabel Caballero \inst{\ref{inst2}}
\and Andrea Santangelo \inst{\ref{inst1}}
}
\institute{Institut f\"ur Astronomie und Astrophysik, Universit\"at T\"ubingen, Sand 1, D-72076 T\"ubingen, Germany,\\
e-mail: daniela.mueller@astro.uni-tuebingen.de\label{inst1} \and
Laboratoire AIM, CEA/IRFU, CNRS/INSU, Universit\'e Paris Diderot, CEA DSM/IRFU/SAp, 91191 Gif-sur-Yvette, France\label{inst2}}
\date{Received /
Accepted}

\abstract {
  A number of accreting X-ray pulsars experience 
  spectral changes, both on the long time scales and on the time scales of 
  the neutron star spin period. The sources seem to form two distinct groups
  which differ by the type of the spectral variations with flux.
  Such a bimodality probably reflects two different regimes of accretion that
  may result in a particular pulsar depending on its luminosity -- 
  so-called sub- and super-critical regimes.
} 
{
  We investigated the behavior of the spectral parameters of the Be/X-ray 
  binary system A\,0535+26, as a function of flux and pulse phase.
}
{
  We used the data collected with \emph{INTEGRAL} and \emph{RXTE} during 
  the April 2010 outburst of the source. We analyzed the 
  phase-averaged and phase-resolved spectra and performed pulse-to-pulse 
  spectral analysis of the pulsar.
}
{
  Our analysis reveals variability in the continuum parameters of
  the source's pulse-averaged spectrum with flux. 
  The pulse-averaged cyclotron line 
  energy does not change with the source luminosity during the outburst,
  which is consistent with previous studies. Our 
  pulse-phase resolved and pulse-to-pulse analyses reveal, however, 
  indications for a positive correlation of the cyclotron line energy with 
  flux, as well as a flux-dependence of the continuum parameters. 
  Based on the observed
  behavior, we argue that A\,0535+26 operates at the sub-critical
  accretion regime. 
}
{}

\maketitle

\section{Introduction}
\label{sec:introduction}
 
The X-ray transient A\,0535+26 \footnote{A\,0535+26 is referred to as 1A\,0535+262 in SIMBAD Astronomical Database} was discovered by the satellite \emph{Ariel\,V} during a giant outburst in April 1975 \citep{1975Natur.256..628R}. 
The source belongs to the class of Be/X-ray binaries. It consists of an
accreting neutron star orbiting the O9.7\,IIIe optical star HDE 245770 
\citep{1979ApJ...228..893L,1980A&AS...40..289G}. The system regularly undergoes 
X-ray outbursts separated by quiescence phases \citep{1991ApJ...369..490M}. 
In Be/X-ray binaries, outbursts are often divided into \emph{normal} and \emph{giant} ones. 
Normal outbursts repeat periodically at times near the periastron passage and are thought to originate from an increase of the accretion rate ($L_\mathrm{max}/L_\mathrm{min} \approx 100$) due to a closer position of the neutron star to its Be star companion \citep{1986ApJ...308..669S}. Giant outbursts show a larger increase in luminosity ($L_\mathrm{max}/L_\mathrm{min} \approx $100--1000) and last several tens of days \citep{1986ApJ...308..669S}. 
Often there is, however, no strict distinction between the two types of outbursts. 
Thus, some outbursts in A\,0535+26 can not be clearly assigned to one of those two groups. Besides the outburst in 1975, A\,0535+26 went through giant outbursts in 1980 \citep{1982ApJ...263..814N}, 1983 \citep{1990ApJ...351..675S}, 1989 \citep{1989IAUC.4769....1M}, 1994 \citep{1996ApJ...459..288F}, 2005 \citep{2005ATel..504....1T}, 2009 \citep{2009ATel.2336....1K, 2009ATel.2337....1C, 2011arXiv1107.3417C} and 2011 \citep{2011ATel.3166....1C,2011ATel.3173....1T,2011ATel.3204....1C}.

The distance of the system has been measured to be around 2\,kpc \citep{1998MNRAS.297L...5S}. Recent measurements of the orbital period give $\sim$111.1\,days \citep{2006HEAD....9.0759F}. The pulse period is around 103.39\,s \citep{2007A&A...465L..21C}. 

The phase-averaged X-ray spectrum of A\,0535+26 can be
described by a hard power law $E^{-\Gamma}$
\citep[$\Gamma$\,=\,0.8--1.1 for 2--18\,keV \emph{Ariel-V/SSI}
  observations,][]{1975Natur.256..631R} modified above
$\sim$15--20~keV by an exponential roll-off
\citep[\emph{TTM/HEXE} spectrum in
  2--156\,keV, ][]{1994A&A...291L..31K}. Alternatively,
  \citet{2011ApJ...733...96A} used the 
  thermal Comptonization model to fit the low-energetic part of the
  X-ray continuum. Additionally, the
spectrum shows absorption lines, interpreted as cyclotron resonant
scattering features (CRSFs).  
Such features are caused by resonant scattering 
of photons off the electrons in Landau levels
\citep[see, e.g.,][]{1978ApJ...219L.105T,1998ApJ...493..154I}.
The energy of the fundamental line and the spacing between
harmonics are directly proportional to the magnetic field strength.
Two cyclotron absorption lines at about 50\,keV and 100\,keV were
reported and interpreted as the fundamental and the first
  harmonic and implying a magnetic field strength of about
$4.3\cdot10^{12}$\,G
\citep{1994A&A...291L..31K,2007A&A...465L..21C}. While the 
  presence of the fundamental line at about 50\,keV was initially questioned
  \citep{1995ApJ...438L..25G,1996A&AS..120C.183A}, newer
  analyses clearly confirm the presence of both lines and their
  interpretation as a fundamental cyclotron feature and its first
  harmonic \citep{2007A&A...465L..21C}.

An evolution of the cyclotron line energy with the source luminosity has been reported for several accreting X-ray pulsars. A negative correlation of the line energy with luminosity has been observed in V\,0332+53 and 4U\,0115+63 \citep{2006MNRAS.371...19T,2007AstL...33..368T,2010MNRAS.401.1628T,2006A&A...451..187M}. A positive correlation was detected for Her\,X-1 and GX\,304-1 \citep{2007A&A...465L..25S,2012A&A...542L..28K}. A similar behavior is seen in the pulse-to-pulse variations. For V\,0332+53 and 4U\,0115+63, a negative correlation of cyclotron line energy with the amplitude of individual pulses and a softening of the spectral continuum with pulse amplitude was reported, whereas for Her\,X-1 and A\,0535+26 a positive correlation of the line energy with pulse amplitude and a hardening of the spectrum with pulse amplitude was observed \citep{2011A&A...532A.126K}. Since A\,0535+26 showed the spectral-flux dependence
on the pulse-to-pulse timescales only, it is important to verify and study this behavior, 
using the data from a new outburst of the source.

Pulse-phase resolved analysis shows that the spectrum of the main peak of the pulse profile 
seems to be harder than the spectrum of the secondary peak \citep{1994A&A...291L..31K,2009PhDT........12C}. \citet{2009PhDT........12C} also found a noticeably
lower cyclotron line energy during the main peak compared to the secondary one. 

In this work, we analyzed the April 2010 outburst of A\,0535+26. On April 3 2010, a brightening of A\,0535+26 in the \emph{Swift/BAT} light curve was reported \citep{2010ATel.2541....1C}. In the 20--100\,keV energy band of \emph{INTEGRAL/ISGRI} the flux of the source reached a value of about 0.92\,Crab and the fundamental cyclotron line was detected at about 47.6\,keV \citep{2010ATel.2541....1C}. The outburst analyzed in this work took place after the giant outburst of December 2009, which reached a flux of about 5.14\,Crab in the 15--50\,keV \emph{Swift/BAT} light curve \citep{2010ATel.2541....1C}. 

Our analysis is focused on the spectral evolution of A\,0535+26 as a function of luminosity. We analyzed phase-averaged, pulse-phase resolved, and pulse-amplitude resolved (pulse-to-pulse) spectra of the 2010 outburst, to 
explore spectral variations on short (down to the pulsar's spin period) and long 
(several days) time scales. A description of the observations and data processing is given in Section \ref{sec:observations_analysis}. In Section \ref{sec:analysis_results}, we describe the analysis of the phase-averaged, phase-resolved and pulse-to-pulse spectra. In Sections \ref{sec:discussion} and \ref{sec:summary_conslusions}, we discuss and summarize our results.

\section{Observations and data processing}
\label{sec:observations_analysis}

For our analysis, we used data from the \emph{JEM-X} and \emph{IBIS} instruments of the \emph{INTEGRAL} satellite (INTErnational Gamma-Ray Astrophysics Laboratory, \citealt{2003A&A...411L...1W}) and from the \emph{PCA} and \emph{HEXTE} instruments onboard Rossi X-ray Timing Explorer (\emph{RXTE}, \citealt{1993A&AS...97..355B}).

The \emph{JEM-X} monitor provides spectra and images with arc-minute resolution in the 3--35\,keV energy band \citep{2003A&A...411L.231L}. The high angular resolution gamma-ray imager \emph{IBIS} consists of two detector layers, \emph{ISGRI} and \emph{PICsIT} \citep{2003A&A...411L.131U}. The \emph{ISGRI} layer, which
we used for our analysis, 
is sensitive at energies from 15\,keV to 1\,MeV \citep{2003A&A...411L.141L}.
The \emph{PCA} instrument onboard \emph{RXTE} consists of a set of five proportional counters and is sensitive to energies 2--60\,keV \citep{2006ApJS..163..401J}. The high-energy detector \emph{HEXTE} operates in the energy range 15--250\,keV \citep{1998ApJ...496..538R}.\par\medskip

\begin{figure}  \centering
\includegraphics[viewport=0cm 0cm 18cm 13cm,scale=0.5,clip]{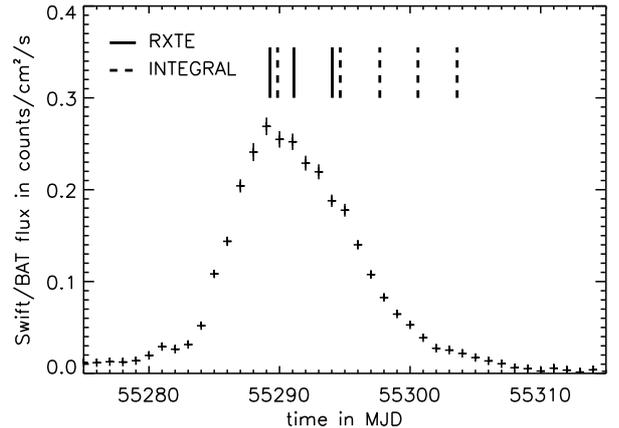}
\caption{The \emph{Swift/BAT} daily light curve of the April 2010 outburst of A\,0535+26. The \emph{INTEGRAL} and \emph{RXTE} observations which we used for our analysis are indicated with dashed (\emph{INTEGRAL}) and solid lines (\emph{RXTE}).}
\label{figure:1}
\end{figure}

We used data covering the time from the maximum to the decay of the April 2010 outburst (2010-04-03 to 2010-04-17, Table~\ref{table:2}). A \emph{Swift/BAT} light curve of A\,0535+26 with the indicated observations is shown 
in Fig.\,\ref{figure:1}.

\begin{table}
\caption{Selected observations.}
\label{table:2}
\centering
\begin{tabular}{c c c} 
\hline\hline
Observation & Exposure [ks] & Flux in Crab [10-100\,keV]\\
\hline
	\emph{INTEGRAL }0912& 65.1 & 0.978 \\
	\emph{INTEGRAL }0914& 23.8 & 0.675 \\
	\emph{INTEGRAL }0915& 28.6 & 0.368 \\
	\emph{INTEGRAL }0916& 29.4 & 0.191 \\
	\emph{INTEGRAL }0917& 31.7 & 0.096 \\ \hline
	\emph{RXTE }95347-02-01-00& 3.3 & 0.908 \\
	\emph{RXTE }95347-02-01-01& 3.3 & 0.986 \\
	\emph{RXTE }95347-02-01-02& 3.0 & 0.715 \\
\hline
\end{tabular}
\end{table}

The \emph{INTEGRAL} data were processed using \emph{OSA}\,9.0. For the \textsl{ISGRI} data we performed an additional gain correction to account for known problems in the \emph{OSA}\,9.0 energy calibration. We used the \textsl{JEM-X} data in the 5--30\,keV range and the 
\textsl{ISGRI} data
in the 20--145\,keV range. We added systematic errors of 3\% for \emph{JEM-X} and 2\% for \emph{ISGRI}, which we estimated by fitting a Crab spectrum accumulated with the respective instruments with a canonical power-law with a photon index of $\sim$2.1. The Crab spectra are taken during the same observations since both sources are present simultaneously in the \emph{INTEGRAL} field-of-view.

We used HEASoft Version 6.11 and 
selected the energy ranges 5--60\,keV for \emph{PCA}
and 20--145\,keV for  
\emph{HEXTE} for the \emph{RXTE} data. 
In the \emph{PCA} spectra we excluded data below 5~keV due to
  an instrumental background feature at the xenon \emph{L} edge around
  $\sim$ 4.8\,keV, which is poorly modelled and might introduce
  systematic effects to our spectral fits \citep{2006ApJ...641..801R,2008A&A...480L..17C}
We added a systematic error of 1\% for \emph{PCA} and 1\% for \emph{HEXTE} data as recommended by the instruments teams. A systematic error for the \emph{HEXTE} data is required due to the stop of the rocking motion of the \emph{HEXTE Cluster A} in 2006 and \emph{HEXTE Cluster B} in December 2009. Since then, \emph{HEXTE Cluster A} is fixed on-source while \emph{HEXTE Cluster B} measures the background, which is then converted to an adequate background for \emph{Cluster A} \citep{2006HEAD....9.1821P}. Similarly to \citet{2011ApJ...733...23R}, we used the \emph{recorn} model in \emph{XSPEC} to account for uncertainties in the correction factor of the background.\\

In order to select time intervals for the phase-resolved and pulse-to-pulse analyses, we extracted light curves for each observation. We then converted the photon arrival times to the solar system barycenter and corrected them for the orbital motion of the binary system using the ephemeris from \citet{2006HEAD....9.0759F}. The pulse period was determined with the epoch-folding technique. 
The pulse phase zero corresponds to the minimum in the pulse profile (see Fig. \ref{figure:pp_phase}).

\begin{table*}
\begin{minipage}[c]{1.0\textwidth}
\caption{Fit parameters for the \emph{INTEGRAL} observations 0912, 0914, 0915, 0916 and 0917 and the \emph{RXTE} observations 95347-02-01-00, 95347-02-01-01 and 95347-02-01-02. The errors are at 1\,$\sigma$ confidence level.}
\label{table:1}
\centering
\renewcommand{\arraystretch}{1.3}
\begin{tabular}{c c c c c c c} 
\hline\hline
Observation & $\Gamma$ & $E_{\mathrm{rolloff}}$ & $E_{\mathrm{cyc}}$ & $\sigma_{\mathrm{cyc}}$ & $\tau_{\mathrm{cyc}}$ & $\chi^2_{\mathrm{red}}$/d.o.f. \\
\hline
	\emph{INTEGRAL }0912&$0.56^{+0.06}_{-0.06}$&$17.21^{+0.44}_{-0.48}$&$44.82^{+0.36}_{-0.37}$&$12.95^{+0.92}_{-0.80}$&$13.16^{+2.08}_{-1.60}$&1.1/110 \\
	\emph{INTEGRAL }0914&$0.30^{+0.11}_{-0.13}$&$15.32^{+0.77}_{-0.90}$&$43.04^{+0.45}_{-0.51}$&$14.67^{+1.51}_{-1.21}$&$24.29^{+6.62}_{-4.37}$&0.9/110 \\
	\emph{INTEGRAL }0915&$0.65^{+0.07}_{-0.07}$&$18.18^{+0.64}_{-0.64}$&$43.99^{+0.41}_{-0.41}$&$10.97^{+0.75}_{-0.68}$&$16.18^{+2.04}_{-1.70}$&0.9/110 \\
	\emph{INTEGRAL }0916&$0.83^{+0.08}_{-0.09}$&$19.88^{+1.00}_{-1.01}$&$44.31^{+0.70}_{-0.67}$&$11.14^{+1.18}_{-1.02}$&$16.82^{+3.24}_{-2.47}$&0.8/110 \\
	\emph{INTEGRAL }0917&$1.01^{+0.08}_{-0.09}$&$23.75^{+1.59}_{-1.45}$&$45.83^{+1.06}_{-0.97}$&$8.67^{+1.22}_{-2.26}$&$14.40^{+2.75}_{-2.26}$&0.9/110 \\ \hline
	\emph{RXTE }95347-02-01-00&$0.44^{+0.02}_{-0.02}$&$16.24^{+0.15}_{-0.15}$&$44.24^{+0.29}_{-0.29}$&$9.01^{+0.36}_{-0.34}$&$8.69^{+0.52}_{-0.49}$&1.2/152 \\
	\emph{RXTE }95347-02-01-01&$0.43^{+0.02}_{-0.02}$&$15.95^{+0.15}_{-0.15}$&$43.92^{+0.29}_{-0.29}$&$9.80^{+0.39}_{-0.34}$&$9.28^{+0.59}_{-0.55}$&1.5/152 \\
	\emph{RXTE }95347-02-01-02&$0.47^{+0.02}_{-0.02}$&$16.26^{+0.18}_{-0.18}$&$43.31^{+0.33}_{-0.33}$&$9.16^{+0.41}_{-0.40}$&$9.00^{+0.63}_{-0.59}$&1.3/152 \\
\hline
\end{tabular}
\end{minipage}
\vspace{0.5cm}

\hspace{0.02\textwidth}
\begin{minipage}[t][][c]{0.43\textwidth}
	\includegraphics[viewport=2.5cm 0.8cm 31cm 32cm,scale=0.28,clip]{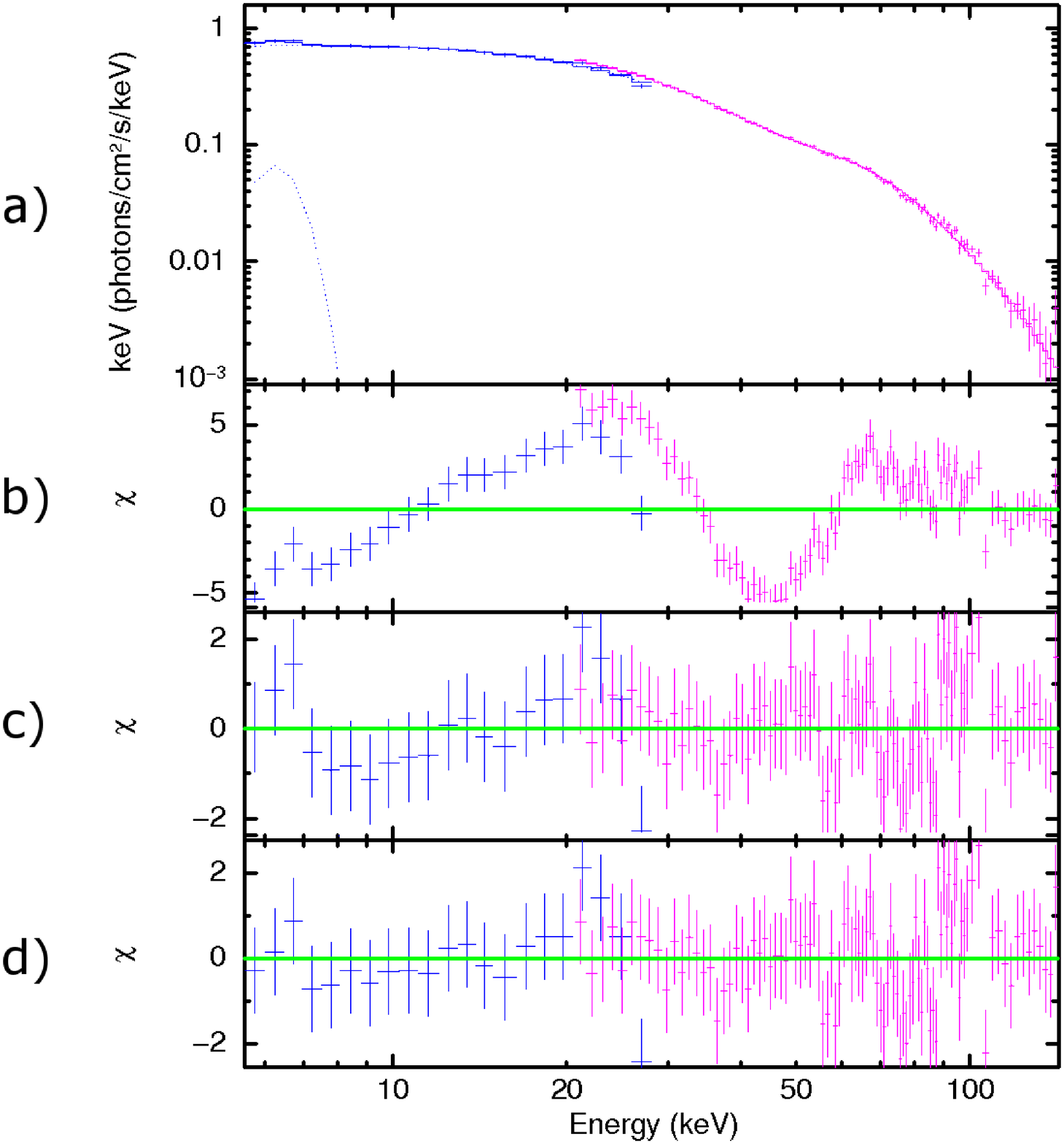}
	\captionof{figure}{The unfolded spectrum of A\,0535+26 from the \emph{INTEGRAL}
  revolution 0912 (a),  the residuals of the fit with the model
  without the 6.4\,keV Fe emission and the cyclotron absorption lines
  (b), the residuals after inclusion in the model of the cyclotron absorption line (c) and the
  Fe emission line (d).}
	\label{figure:spectra}
	\vspace{0.5 cm}
\end{minipage} \hspace{0.06\textwidth}
\begin{minipage}[t][][c]{0.43\textwidth} 
	\includegraphics[width=\textwidth,viewport=0cm 0.5cm 13.5cm 15cm,scale=0.65,clip]{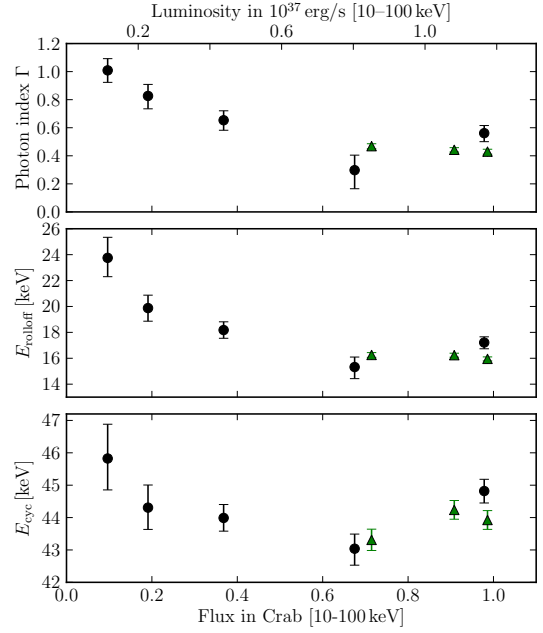}
	\captionof{figure}{Photon index $\Gamma$, rolloff energy $E_\mathrm{rolloff}$ and cyclotron line energy $E_\mathrm{cyc}$ as a function of the source flux in the 
	10--100\,keV band in units of Crab and the luminosity in erg/s 
	for a distance of 2\,kpc (upper X-axis). 
	The \emph{INTEGRAL} observations are marked with circles, the \emph{RXTE} 
	observations -- with triangles. The errors are at 1$\sigma$ confidence
	level.}
	\label{figure:2}
\end{minipage}

\hspace{0.02\textwidth}
\begin{minipage}[t]{0.43\textwidth} 
	\includegraphics[viewport=1cm 0.5cm 25cm 19cm,scale=0.3,clip]{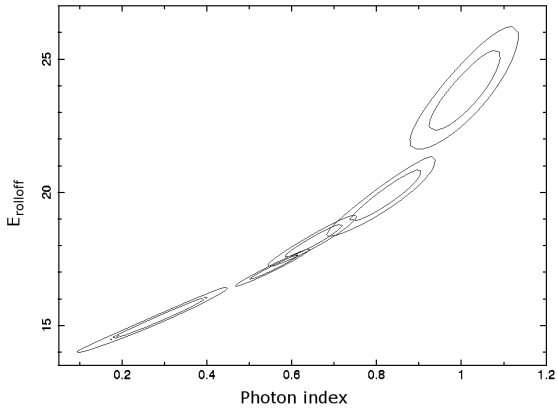}
	\captionof{figure}{A contour plot of the folding energy $E_{\mathrm{rolloff}}$ and the photon index $\Gamma$ for all five \emph{INTEGRAL} observations. The contour levels are 	at $\chi^2_\mathrm{min}+1.0$ (corresponds to 68\% confidence for one chosen parameters) and $\chi^2_\mathrm{min}+2.3$ (corresponds to 68\% confidence for two chosen parameters).}
	\label{figure:contour}
\end{minipage} \hspace{0.06\textwidth}
\begin{minipage}[t]{0.43\textwidth} 
	\includegraphics[width=\textwidth,viewport=0cm 0cm 19cm 12cm,scale=0.45,clip]{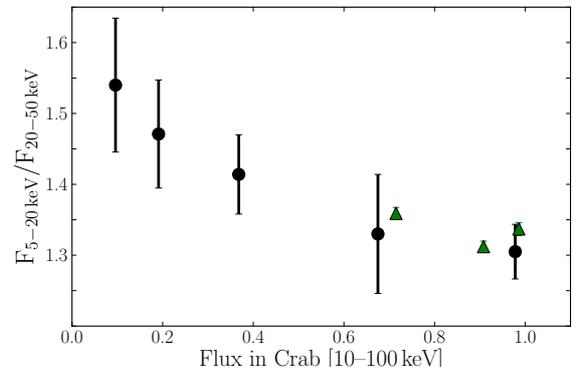}
	\captionof{figure}{The 5--20\,keV to 20--50\,keV flux ratio as a 
	function of the source flux in units of Crab. 
	The graph indicates a hardening the spectrum with flux.
	}
	\label{figure:hardness_ratio}
	\end{minipage}
\end{table*}

\section{Spectral analysis}
\label{sec:analysis_results}

\subsection{Phase averaged spectra}
\label{sec:phase_averaged_spectra}

The spectra of A\,0535+26 for various luminosity states were best fitted by a 
power-law with a photon index $\Gamma$ modified
by an exponential rolloff with the folding energy $E_{\mathrm{rolloff}}$ (the \texttt{cutoffpl} model in \emph{XSPEC}): \\

$F(E) = \mathrm{const} \cdot E^ {- \Gamma } \cdot e^ {-E/E_{\mathrm{rolloff}}}. $ \\

We included an emission line at around 6.4\,keV to model the Fe K$\alpha$ 
fluorescence line. Additionally, 
the residuals of the fit require inclusion of a 
an absorption line around 40\,keV (Fig. \ref{figure:spectra}) which we
interpret as a CRSF (see Introduction). To account for this feature, 
we included a multiplicative absorption line with a Gaussian optical 
depth profile (the \emph{XSPEC} \texttt{gabs} model) with a centroid energy of
$\sim$44.4\,keV:

\begin{equation}
F(E) = \exp \left[-\frac{\tau_\mathrm{cyc}}{\sqrt{2\pi}\sigma_\mathrm{cyc}}\exp \left(-\frac{(E-E_\mathrm{cyc})^2}{2\sigma_\mathrm{cyc}^2}\right)\right],
\end{equation}

where $E_\mathrm{cyc}$ is the cyclotron line centroid energy, 
$\tau_\mathrm{cyc}$ and $\sigma_\mathrm{cyc}$ characterize the line 
strength and width respectively. 
The \emph{RXTE} observations also indicate the presence of the first harmonic
 line around 104\,keV. The best-fit parameters for the \emph{INTEGRAL} and 
\emph{RXTE} observations are summarized in Table~\ref{table:1}. 
Figure~\ref{figure:2} shows the photon index $\Gamma$, the rolloff 
energy $E_{\mathrm{rolloff}}$ and the cyclotron line energy $E_\mathrm{cyc}$ 
as a function of 10--100\,keV flux in Crab units. The latter is obtained
from a direct comparison of the source flux with that of the
simultaneously observed Crab nebula.
Because the first harmonic in the \emph{RXTE} spectra is not statistically
significant and to be consistent with the \emph{INTEGRAL} spectral 
model, only one Gaussian absorption line was included in our spectral model, accounting for the fundamental cyclotron line.

The photon index decreases from $\sim$1.0 for low luminosities to 
$\sim$0.5 for the outburst maximum, indicating a hardening of the power-law 
continuum with flux. The rolloff energy $E_{\rm rolloff}$, however, decreases 
from $\sim$24 to $\sim$16\,keV from the lowest to highest luminosity.
That is, the part of the continuum around $E_{\rm rolloff}$ softens with
flux.
No correlation of the cyclotron line energy with the source luminosity 
is observed. The cyclotron line energy, however, varies irregularly 
between $\sim$43\,keV and $\sim$46\,keV. To rule out any coupling between 
$E_{\rm rolloff}$  and $\Gamma$, we produced contour plots of those 
two parameters (Fig. \ref{figure:contour}). Since the evolution of $\Gamma$ 
indicates a hardening of the spectrum with increasing flux while the
variation of $E_{\rm rolloff}$ suggests the opposite behavior, 
we calculated the hardness ratio to study the spectral variations
independently of the spectral model. Figure \ref{figure:hardness_ratio} shows 
a ratio of the 5--20\,keV flux to the 20--50\,keV flux as a function of the
10--100\,keV flux. The observed decrease of the ratio implies an overall 
hardening of the spectrum with flux. A similar behavior is reported in \citet{2013ApJ...764...L23} for the normal double-peaked outburst of August 2009. The results of this analysis, i.e. no correlation for the cyclotron line energy and a spectral hardening with flux, seem to support our results in the analyzed luminosity range.
Following \citet{2011ApJ...733...96A}, we tried to model the
  spectra with the \texttt{compTT} model in \emph{XSPEC}. We find that
  although the \texttt{compTT} model provides a good description of
  the X-ray spectrum below $\sim$35\,keV (the energy range used in
  \citealt{2011ApJ...733...96A}),
  it produces strong residuals when applied to the overall energy range evaluated in this paper, thus making the \texttt{compTT} overall fit formally unacceptable. Nevertheless, we
  analyzed the evolution of the cyclotron line energy as a function of
  flux using \texttt{compTT} and found it to be consistent with the
  results, obtained with our phenomenological model. 

\begin{figure*} \centering
\includegraphics[viewport=0.5cm 0cm 14cm 15cm,scale=0.54,clip]{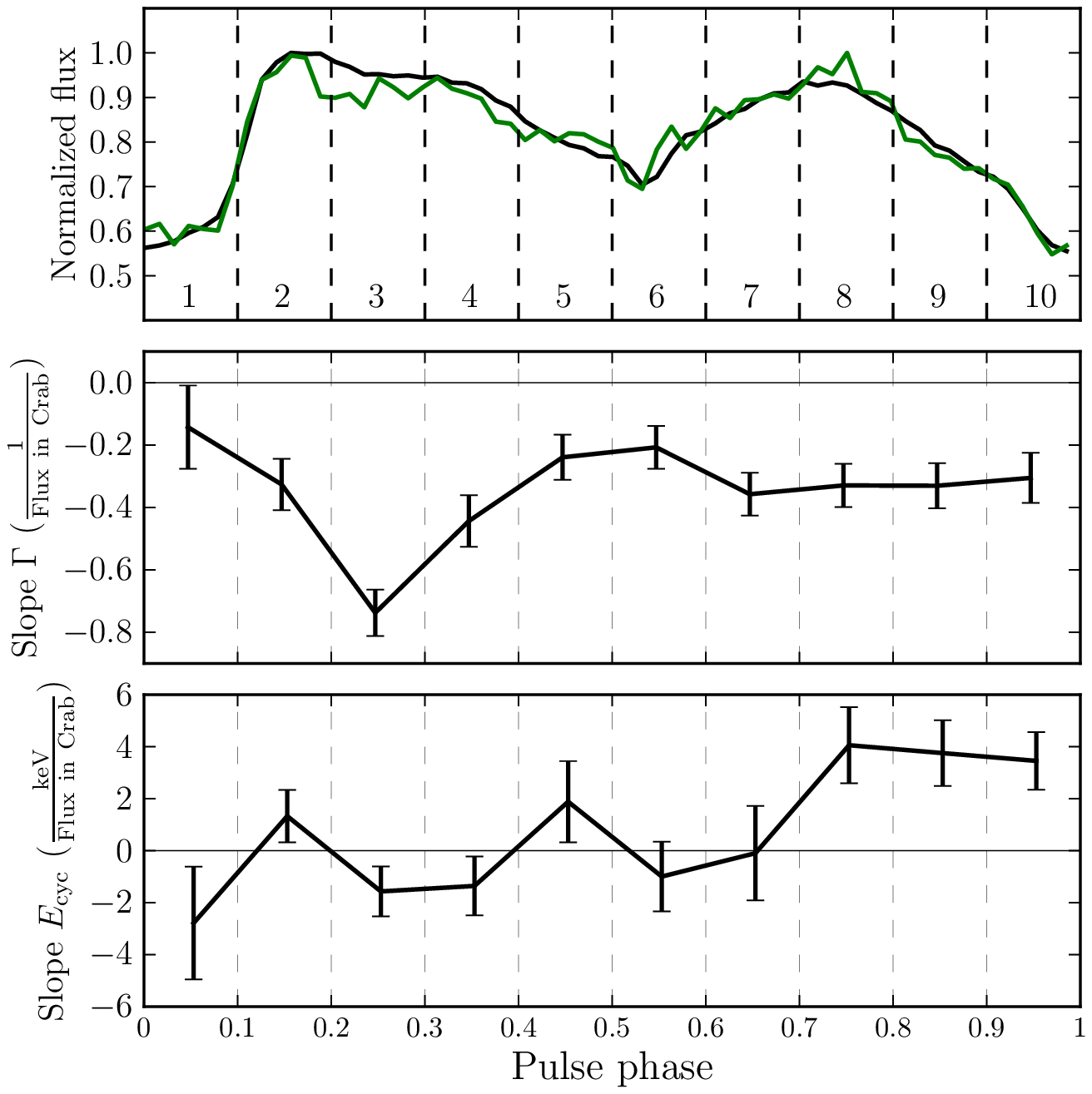}
\includegraphics[viewport=0.55cm 0.7cm 13.8cm 24cm,scale=0.4,clip]{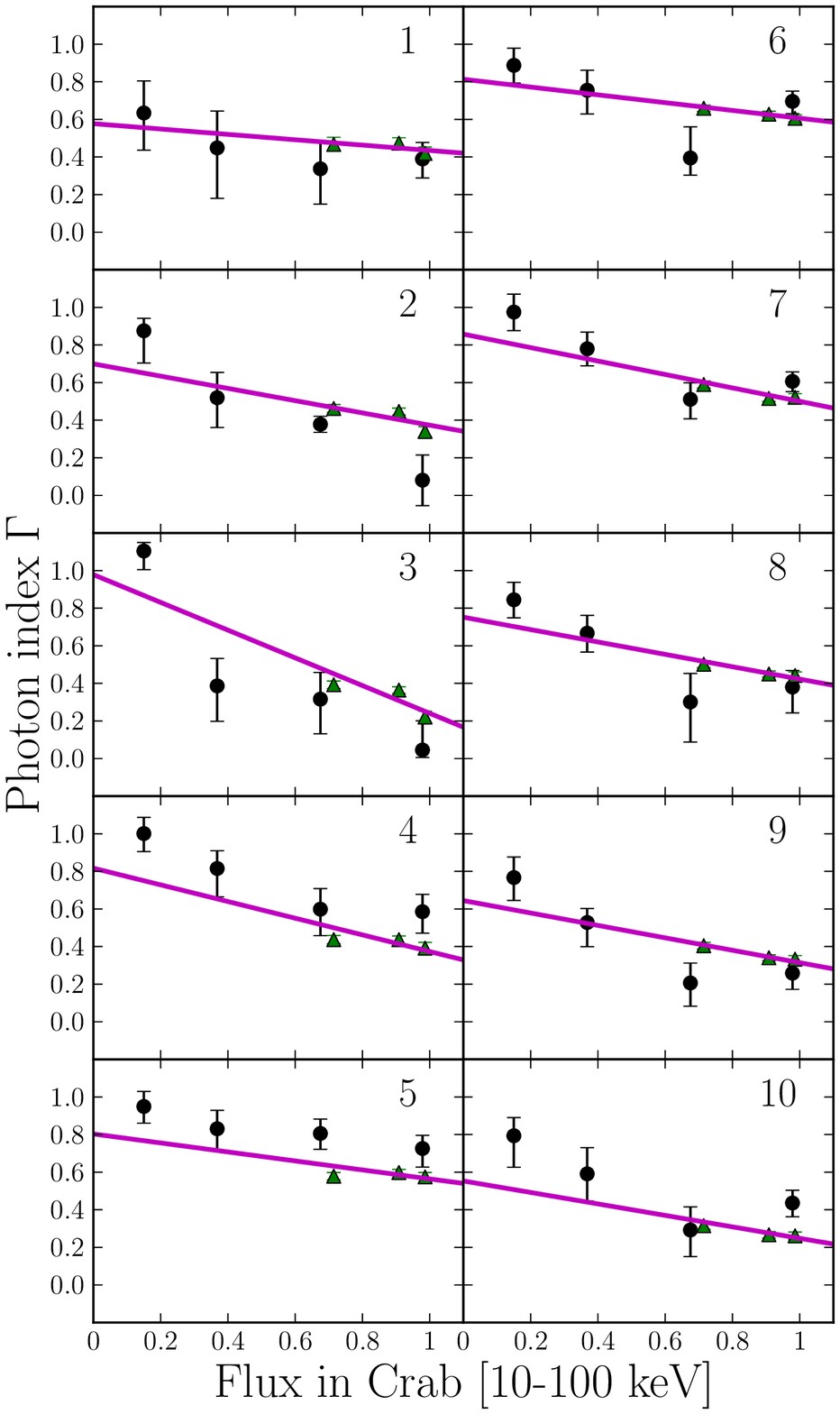}
\includegraphics[viewport=0.55cm 0.7cm 13.7cm 24cm,scale=0.4,clip]{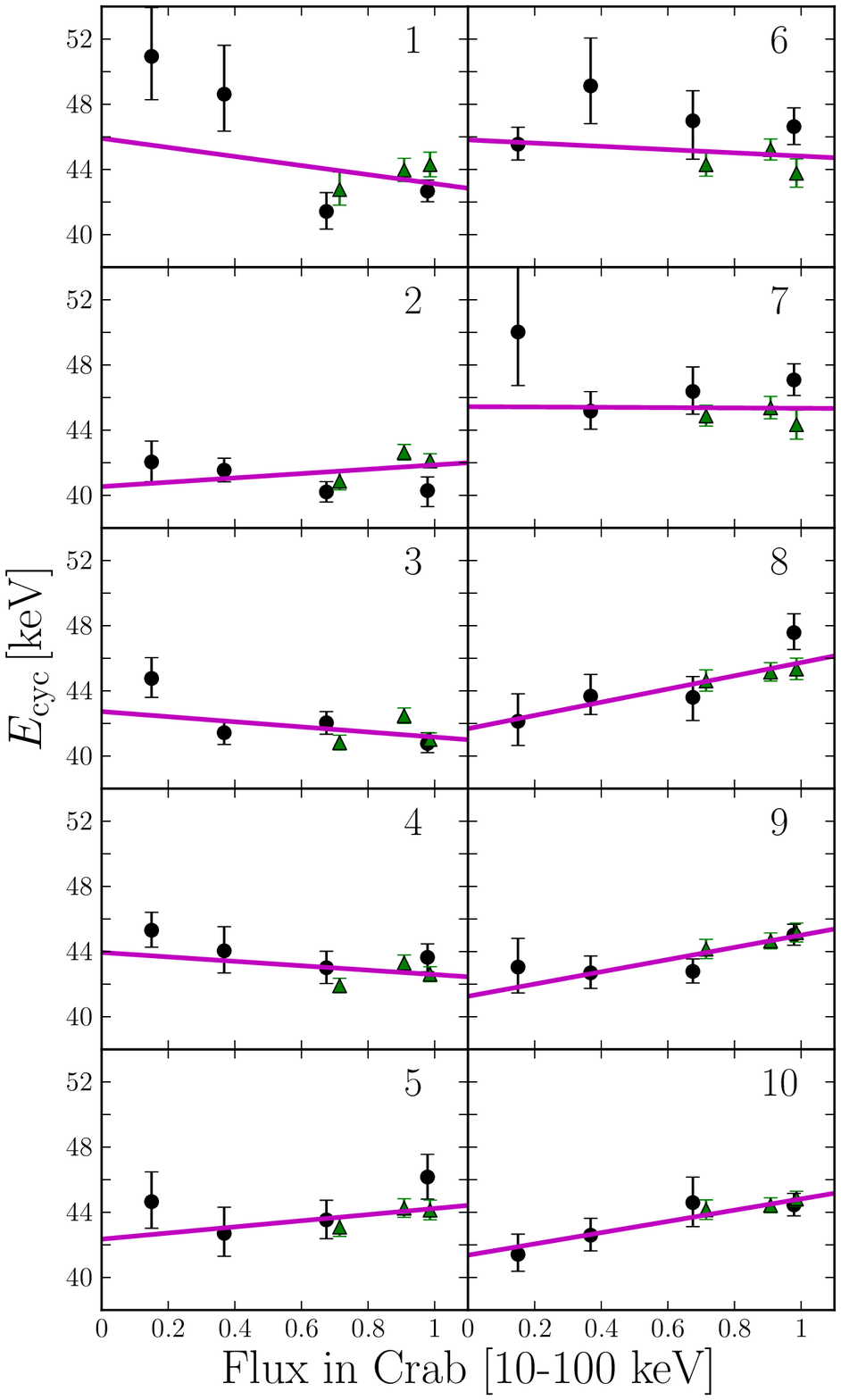}
\caption{The best-fit parameters of the pulse-phase resolved
    spectra at different flux levels with 1\,$\sigma$
    errors. \emph{Left column:} In the top panel, the averaged and
    normalized pulse profile of the \emph{INTEGRAL} observation during
    rev. 0912 (3--80\,keV, black line) and the \emph{RXTE} observation
    95347-02-01-01 (3--60\,keV, green line) are shown. The pulse-phase
    intervals for which the spectra were extracted are indicated by
    vertical dashed lines and numbered from ``1'' to ``10''. Within
    each bin, we analyzed the spectral parameters as a function of
    flux. The middle and bottom panels represent
    the slopes with associated 1$\sigma$-uncertainties obtained from
    the linear fits to the $\Gamma$-flux and 
    $E_{\rm cyc}$-flux dependencies in each pulse-phase bin which are
    shown in the middle and right columns. \emph{Middle and right columns:}
    The photon index $\Gamma$ and the cyclotron line energy 
    $E_{\rm cyc}$ as a function of the 
    10--100\,keV flux for each of the ten phase intervals
    marked by the numbers. The
    solid lines show the linear fits to the data points. The \emph{RXTE} 
    observations are marked with triangles, while the \emph{INTEGRAL}
    observations -- with circles.} 
\label{figure:6}
\end{figure*}

\subsection{Pulse-phase resolved spectroscopy}
\label{sec:pulse_resolved_spectra}

To select pulse-phase intervals, we produced pulse profiles which are 
similar to those reported earlier (e.g. \citealt{2009PhDT........12C} and \citealt{2011A&A...526A.131C}) and do not show strong variation with flux. We then extracted spectra in 10 equal phase bins for each observation. That is, we performed pulse-phase resolved spectroscopy separately for different flux levels. Figure \ref{figure:6} summarizes the evolution of the photon index and the 
cyclotron line energy with pulse phase at different fluxes. For each pulse-phase bin 
separated by vertical dashed lines and marked by a number from 1 to 10, \emph{left} column, $\Gamma$ and $E_{\mathrm{cyc}}$ are shown as a function of
10--100\,keV flux in the \emph{middle} and the \emph{right} column, respectively. To increase the photon statistics in the observations
at the lowest flux levels, we summed up the observations from the 
\emph{INTEGRAL} revolutions 0916 and 0917. 
For each pulse-phase bin, we performed a linear fit of the $\Gamma$--flux 
and $E_\mathrm{cyc}$--flux dependencies shown by the solid lines. 
The slopes of the linear fits with the corresponding 1$\sigma$ uncertainties 
are shown in the bottom panels of the \emph{left} column. The value of the slope with the associated uncertainty gives a rough estimate of the significance of the linear correlation between the parameters.

\begin{figure}
\includegraphics[viewport=0cm 5.5cm 10.5cm 17.5cm,scale=0.82,clip]{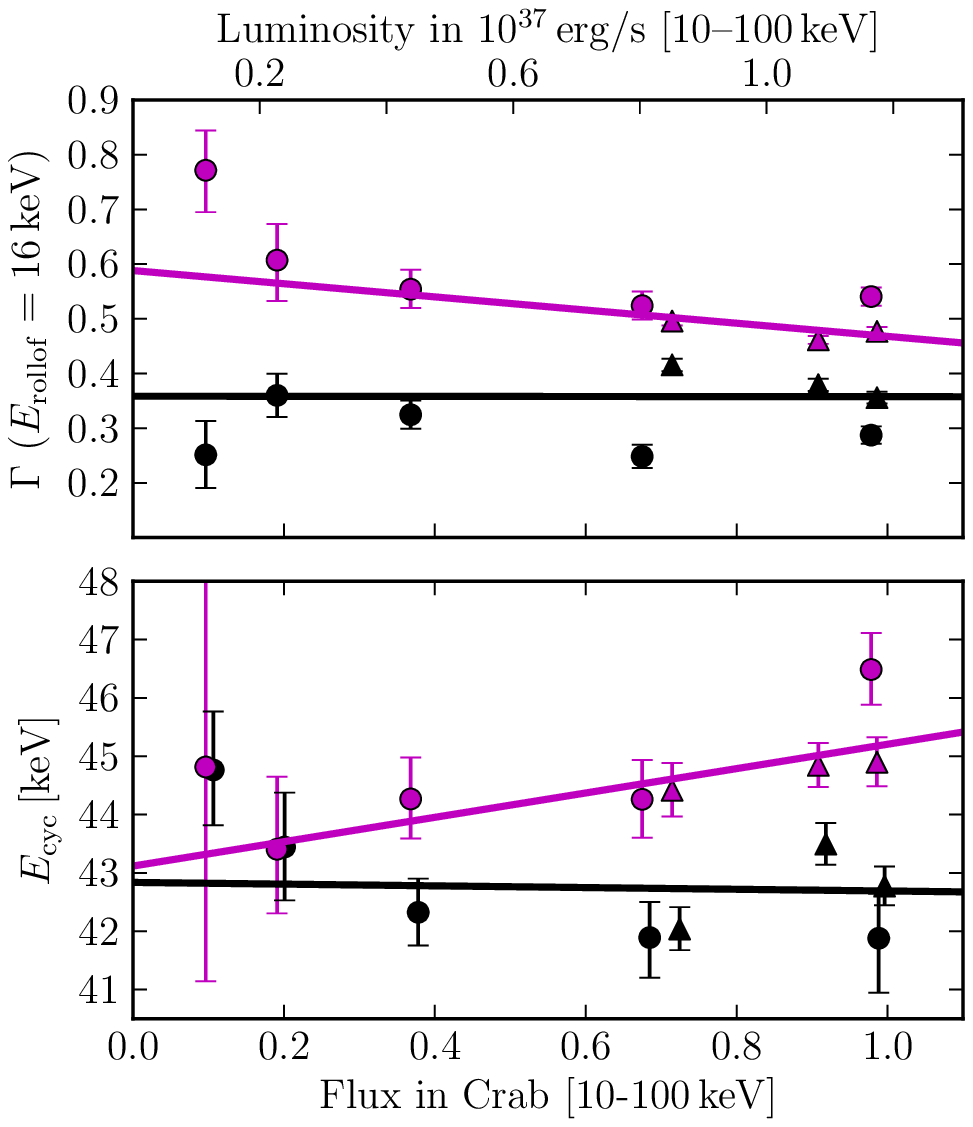}
\caption{Best-fit spectral parameters as a function of flux for the main peak
(phase interval 0.1--0.5) and the secondary peak (phase interval 0.6--1.0) of 
the pulse profile. The data points and the linear fit (solid line) 
corresponding to the main peak are shown in black. The data corresponding
to the secondary peak are shown in magenta. As in previous figures, the 
\emph{INTEGRAL} observations are marked with circles, the \emph{RXTE} 
observations -- with triangles.
For the upper plot, $E_\mathrm{rolloff}$ was fixed to 16.0\,keV (see text). 
For the bottom plot all fit parameters are kept free. The corresponding values for the linear fit and the significance are shown in Table~\ref{table:pulse_phase}.}
\label{figure:phase_01_05_06_10_fit}
\end{figure}

$\Gamma$ and $E_\mathrm{cyc}$ show systematic variations 
with pulse phase and flux. The evolution of $\Gamma$ as a function of
flux, for individual  
pulse-phase intervals, indicates a hardening of the
power-law part of 
the continuum with flux in almost all phase bins. In the pulse phases between 
0.7 and 1.0 there is a strong indication (slope/error
  $\geq$3\,$\sigma$) for a \emph{positive} correlation  
of the cyclotron line energy $E_{\rm cyc}$ with flux (bottom
  panel of \emph{left} column in Fig. \ref{figure:6}). To
  estimate quantitatively the significance of the non-zero slopes,
  we performed a F-test, which yields a value $P$ of the probability that the
  improvement in $\chi^2$ obtained by moving from a fit of the
  dependences by a constant to a
  linear fit with non-zero slope is due to statistical
  flucuations. Lower values of $P$, thus, indicate larger significance.
  In case of the $\Gamma$--flux dependence, the $P$-values for most
  phase bins are at the level of a few per cent indicating marginal
  significance for individual bins.
  For the cyclotron line energy dependence on flux in the phase interval
  0.7--1.0., i.e. the phase bins 8, 9, and 10, the $P$-values are
  0.027, 0.009, and 0.001, respectively.

To check possible influence of any systematic differences in
  calibration between the \emph{INTEGRAL} and \emph{RXTE}
  instruments on our results, we re-calculated the significances using
  only the \emph{INTEGRAL} data points. For the dependence of $\Gamma$
  on flux, the $P$-values remained at a level of a few per cent. For
  the $E_{\rm}$--flux dependence in the phase bins 8, 9, and 10, the
  $P$-values are found to be 0.08, 0.17, and 0.05, respectively. The
  values for the slope, however, did not change considerably. We find,
  therefore, no indication that the measured correlations are driven
  by systematic differences between the instruments onboard the
  two satellites.

As before, to check the stability of our results with respect
  to the choice of the continuum model, following
  \citet{2011ApJ...733...96A} we fitted the spectra using the
  thermal Comptonization model \texttt{compTT} in \emph{XSPEC}. Even
  though the spectral fits are formally unacceptable (similar to the case of
  phase-averaged analysis, see Section 3.1), the evolution of
  the cyclotron line energy with flux is well reproduced. 

In the following, the two peaks of the double-peaked pulse profile are referred to as the main and secondary pulse. To characterize spectral variations within larger pulse-phase intervals,
namely in the main and secondary pulses, we extracted spectra for two 
broad pulse-phase intervals for each observation. The first interval, 0.1--0.5, contains the main 
peak, the second one, 0.6--1.0, -- the secondary peak. The best-fit parameters,
as a function of flux, for the two phase intervals are shown in 
Fig.\,\ref{figure:phase_01_05_06_10_fit}.
To characterize the hardness of the continuum by the photon index 
$\Gamma$ only, we fixed the rolloff energy $E_{\rm rolloff}$ to its averaged value of 16\,keV for the upper plot
of Fig. \ref{figure:phase_01_05_06_10_fit}. The spectrum is harder and the cyclotron line energy lower for the main pulse. We observe an indication for a 
spectral hardening and a \emph{positive} trend of the cyclotron line energy 
with flux for the secondary peak. The values for the slopes of the linear fit 
together with the results of the linear correlation analysis are shown in Table~\ref{table:pulse_phase}.

\begin{table*}
\caption{The slopes of the linear fits to the dependence of the spectral parameters on flux with the corresponding 
Pearson correlation coefficients 
and the associated two-sided null-hypotheses probabilities (a lower value 
indicates higher significance of the correlation). 
The uncertainties for the slopes are at 1$\sigma$ confidence level.
The parameters from the 
pulse-phase resolved analysis of the phase intervals 0.1--0.5 (main peak) 
and 0.6--1.0 (secondary peak) are shown in the top section 
(for $\Gamma$, $E_\mathrm{rolloff}$ was fixed to 16\,keV). 
In the bottom section, the values from the pulse-to-pulse analysis are shown. 
For the fit of $\Gamma$ we fixed $E_\mathrm{rolloff}=15.5$\,keV for 
\emph{INTEGRAL} and $E_\mathrm{rolloff}=15.2$\,keV for \emph{RXTE}. 
For comparison, the values from Klochkov et al. (2011) are also shown.}
\label{table:pulse_phase}
\centering
\renewcommand{\arraystretch}{1.3}
\begin{tabular}{l l l l l}
\hline\hline
Observation & parameter & linear fit & correlation coeff. & probability \\ \hline
\textbf{Pulse-phase-resolved analysis} \\
phase 0.1--0.5 & $\Gamma$  & $-0.00\pm0.03$ (Crab unit)$^{-1}$ & $\sim$0.28 & $\sim$0.51 \\
& $E_\mathrm{cyc}$ & $-0.15\pm0.70$\,keV (Crab unit)$^{-1}$ & $\sim$-0.57 & $\sim$0.14 \\
phase 0.6--1.0 & $\Gamma$ & $-0.12\pm0.03$ (Crab unit)$^{-1}$ & $\sim$-0.83 & $\sim$0.01 \\
& $E_\mathrm{cyc}$ & $2.09\pm0.95$\,keV (Crab unit)$^{-1}$ & $\sim$0.60 & $\sim$0.12 \\ \hline \hline
\textbf{Pulse-to-pulse analysis} \\
\emph{INTEGRAL} & $\Gamma$ & (-5.79 $\pm$ 1.66) $\times$ $10^{-4}$\,/(ISGRI cts/s) & $\sim$-0.96 & $\sim$0.01 \\
& $E_\mathrm{cyc}$ & (0.69 $\pm$ 6.78) $\times$ $10^{-3}$\,keV/(ISGRI cts/s) & $\sim$-0.17 & $\sim$0.78 \\
\emph{RXTE} & $\Gamma$ & (-9.12 $\pm$ 1.84) $\times$ $10^{-5}$\,/(PCA cts/s) & $\sim$-0.99 & $\sim6.89\times10^{-4}$ \\
& $E_\mathrm{cyc}$ & (2.02 $\pm$ 1.26) $\times$ $10^{-3}$\,keV/(PCA cts/s) & $\sim$0.78 & $\sim$0.12 \\ 
\textbf{values from Klochkov et al. (2011)}\\
\emph{INTEGRAL} & $\Gamma$ & (-8.71 $\pm$ 0.86) $\times$ $10^{-2}$\,/(ISGRI cts/s) & -0.96 & 0.01 \\
& $E_\mathrm{cyc}$ & (1.36 $\pm$ 0.53) $\times$ $10^{-2}$\,keV/(ISGRI cts/s) & +0.94 & 0.02 \\
\emph{RXTE} & $\Gamma$ & (-1.54 $\pm$ 0.19) $\times$ $10^{-4}$\,/(PCA cts/s) & -0.98 & $4\times10^{-3}$ \\
& $E_\mathrm{cyc}$ & (1.92 $\pm$ 0.92) $\times$ $10^{-3}$\,keV/(PCA cts/s) & +0.70 & 0.19 \\ \hline 

\end{tabular}
\end{table*}

\subsection{Pulse-to-pulse analysis}
\label{sec:pulse_to_pulse_analysis}

\begin{figure*}
\begin{minipage}[c]{0.47\textwidth}
	\includegraphics[width=\textwidth,viewport=0cm 7cm 13cm 12cm,scale=0.7,clip]{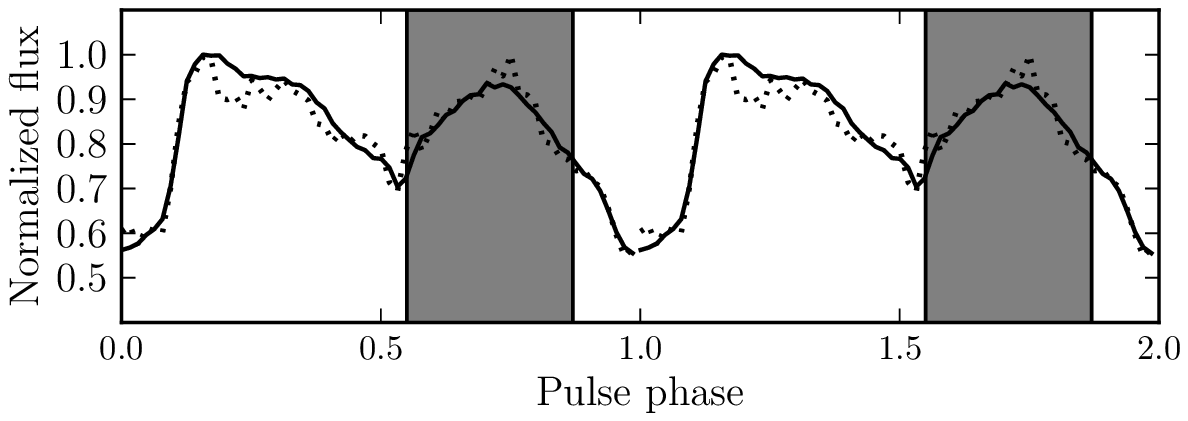}
	\caption{The averaged pulse profile accumulated with \emph{INTEGRAL} 
during its revolution 0912 (3--80\,keV, continuous line) and with \emph{RXTE} 
in the observation 95347-02-01-01 (3--60\,keV, dotted line). The shaded
area indicates the pulse-phase interval used for the pulse-to-pulse analysis.}
	\label{figure:pp_phase}
	\vspace{-0.25cm}
\end{minipage}
\begin{minipage}[c]{0.47\textwidth}
	\includegraphics[width=\textwidth,viewport=0cm 0cm 15cm 9cm,scale=0.63,clip]{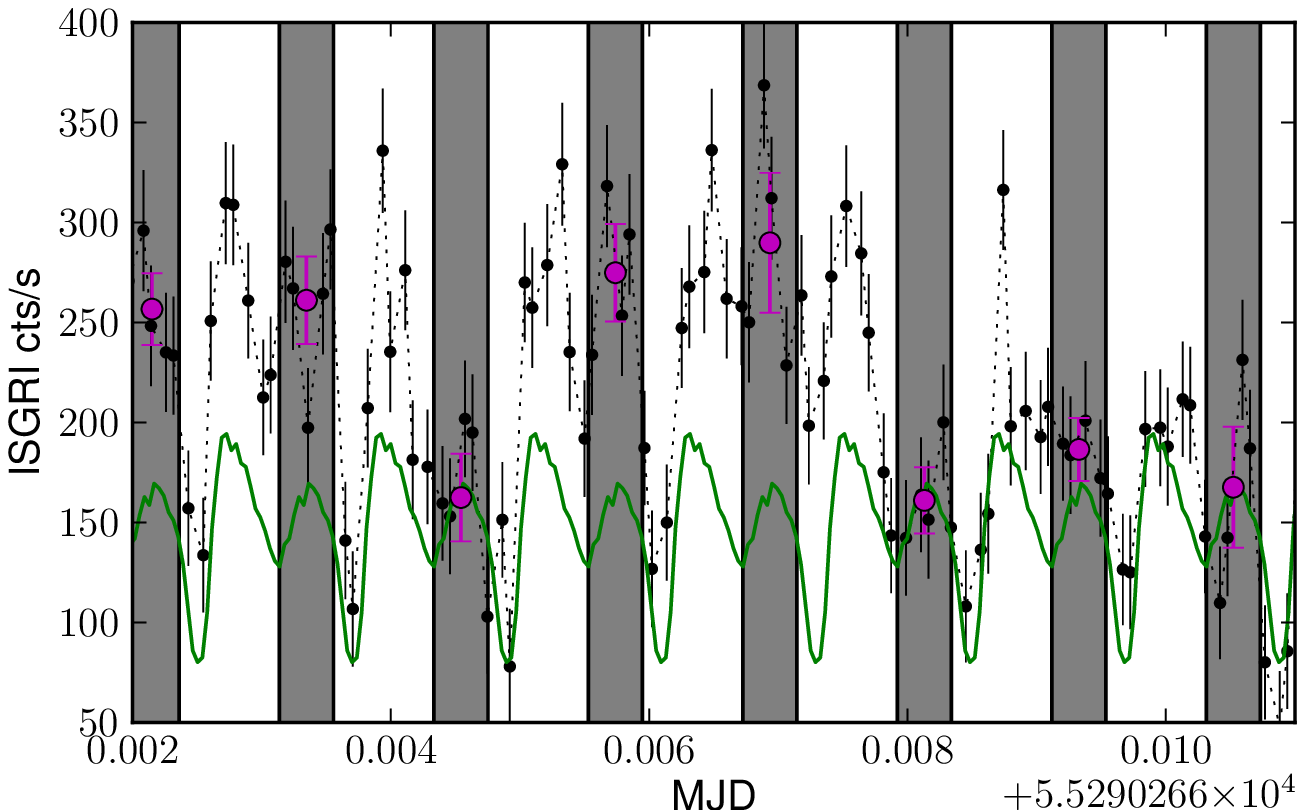}
	\caption{The 20--120\,keV \emph{ISGRI} light curve of the 
\emph{INTEGRAL} observation during rev. 0912
(black data points) together with the overplotted repeated averaged 
pulse profile (solid curve). The selected pulse-phase interval is indicated
by shaded areas. The pulse-to-pulse variability can clearly be seen. 
The measured amplitudes of individual pulses (see text) are shown with 
magenta circles.}
	\label{figure:0912_lc}
	\vspace{1 cm}
\end{minipage}
\begin{minipage}[c]{1.0\textwidth} \centering
	\includegraphics[viewport=0cm 0cm 13cm 10cm,scale=0.55,clip]{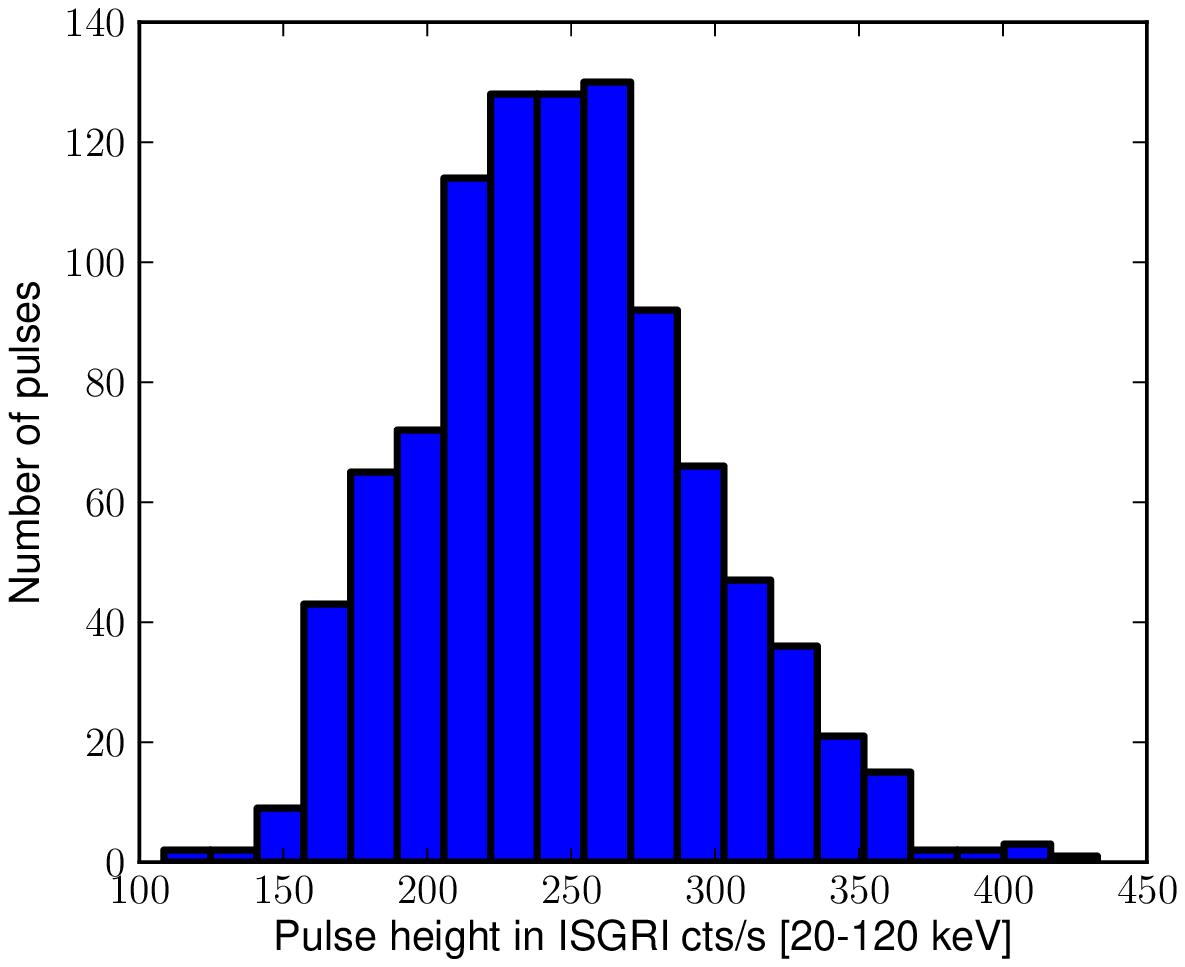} \hspace{1.5cm}
	\includegraphics[viewport=0cm 0cm 13cm 10cm,scale=0.55,clip]{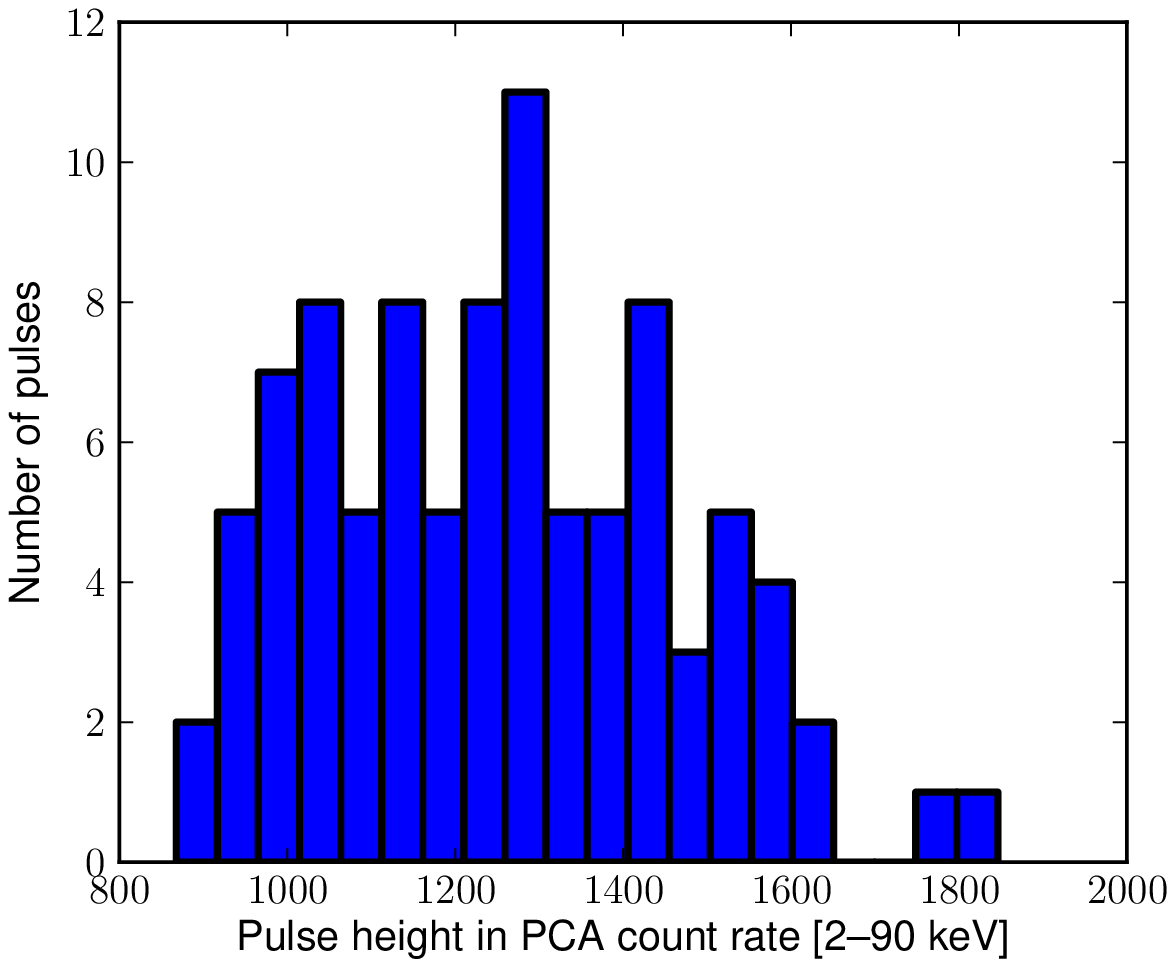}
	\caption[c]{Distributions of individual pulse heights measured with 
\emph{INTEGRAL} (left) and \emph{RXTE} (right).}
	\label{figure:0912_pulse_height_distribution}
	\vspace{1 cm}
\end{minipage}
\begin{minipage}[c]{1.0\textwidth}
\includegraphics[viewport=0cm 0cm 13cm 12cm,scale=0.67,clip]{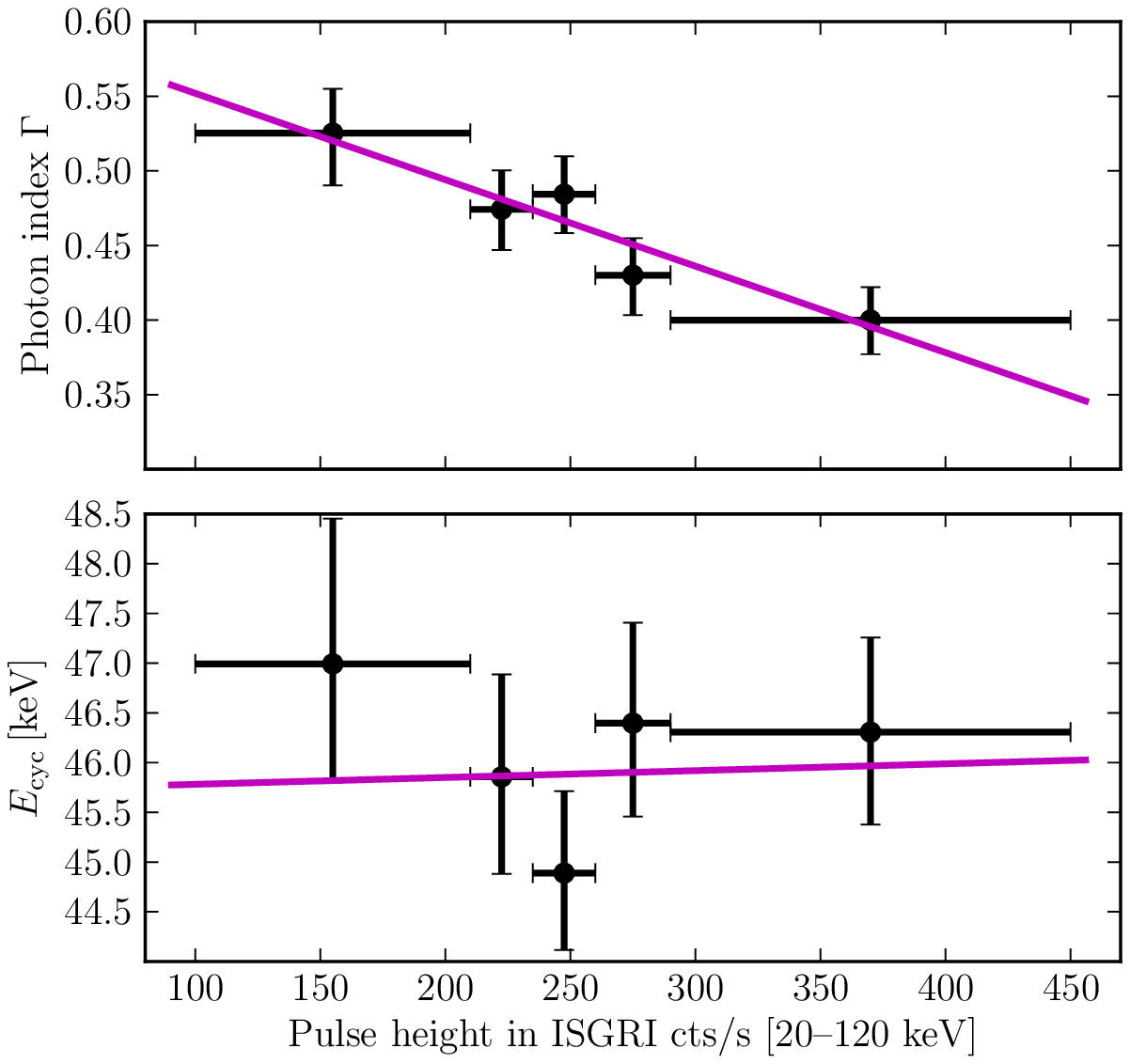}
\includegraphics[viewport=0cm 0cm 13cm 12cm,scale=0.67,clip]{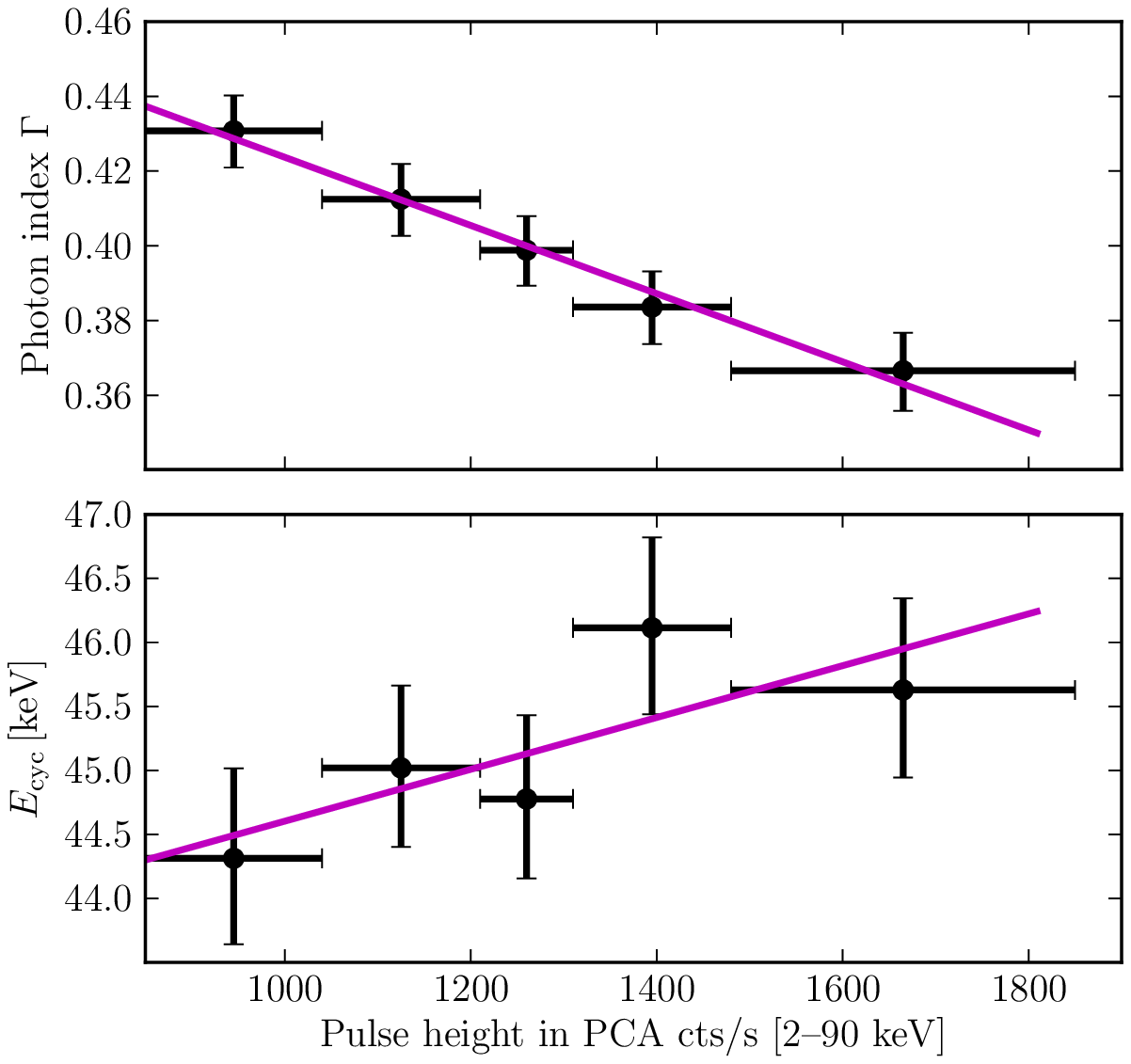}
\caption{\emph{Left:} The results of the pulse-to-pulse analysis of the 
\emph{INTEGRAL} observation of rev. 0912. The photon index 
$\Gamma$ with fixed $E_{\mathrm{rolloff}}$=15.5\,keV and the cyclotron line 
energy $E_{\mathrm{cyc}}$ are plotted as a function of the individual pulse 
height. The vertical error bars indicate 1$\sigma$-uncertainties.
The horizontal error bars correspond to the pulse-height bins used to 
accumulate the spectra. The solid lines indicate linear fits to the data. \emph{Right:} The same but for the \emph{RXTE} 
observations. In the data for the upper panel $E_{\mathrm{rolloff}}$ was
fixes to 15.2\,keV.}
\label{figure:12}
\end{minipage}
\end{figure*}

We performed a spectral analysis of the source on a pulse-to-pulse basis, 
similar to \citet{2011A&A...532A.126K}. To identify individual pulsations,
we extracted light curves for all \emph{INTEGRAL} and \emph{RXTE} observations.
Following \citet{2011A&A...532A.126K}, we have chosen the
  secondary peak of the pulse profile  (Fig. \ref{figure:pp_phase})
  for our analysis, to be able to compare our results. In the pulse
  profiles above $\sim 20--30$\,keV the secondary peak is higher than
  the first one, thus, providing better statistics for the
  measurements of the cyclotron line around 50\,keV. 
For each \emph{individual} pulsation cycle,
we calculated the mean \emph{ISGRI} (for the \emph{INTEGRAL} observations) 
and \emph{PCA} (for the \emph{RXTE} observations) count rate in the selected
pulse-phase interval, which we refer to as \emph{height} or \emph{amplitude}
of individual \emph{pulses}. We then grouped pulses of a similar 
pulse height together to accumulate spectra as a function of pulse height.

For the pulse-to-pulse analysis of the \emph{INTEGRAL} observations, 
we only used the data from the revolution 0912 taken during the maximum of 
the outburst, which provided the best photon statistics. 
Strong pulse-to-pulse variability can 
be seen clearly in Fig. \ref{figure:0912_lc}, where the 20--120\,keV 
\emph{ISGRI} light curve, together with the repeated averaged pulse profile 
and the selected pulse phase interval, is shown. The distribution of the
measured amplitudes of individual pulses
for this observation is shown in the left panel of
Fig.\,\ref{figure:0912_pulse_height_distribution}. We accumulated spectra in
five pulse-height bins: 100--210, 210--235, 235--260, 260--290, and 290--450 ISGRI counts/s.

For \emph{RXTE}, we analyzed all three observations together as they were
performed at a similar luminosity levels, close to the maximum of the 
source's outburst. The selected pulse height intervals for which the spectra have been accumulated are 850--1040, 1040--1210, 1210--1310, 1310--1480, and 
1480--1850 PCA counts/s. The distribution of pulse heights for the 
\emph{RXTE} data is shown in the right panel of Fig.\,\ref{figure:0912_pulse_height_distribution}.

The \emph{INTEGRAL} and \emph{RXTE} observations were analyzed separately 
because of possible systematic effects of non-perfect cross-calibration 
between the two observatories. The best-fit photon index and cyclotron line energy as a function of pulse height are shown in Fig.\,\ref{figure:12}. We fixed the rolloff energy $E_\mathrm{rolloff}$ to 15.5\,keV for the \emph{INTEGRAL} and 
15.2\,keV, for the \emph{RXTE} data for the plot of the photon index $\Gamma$. A significant decrease of the photon index (hardening of the power-law
continuum) with pulse height is indicated by the data from both
satellites. The cyclotron line energy, however, does not show any
considerable evolution with pulse amplitude in the \emph{INTEGRAL} data
and an indication of a positive correlation with pulse height in the
\emph{RXTE} data. The results of the corresponding formal 
linear correlation analysis 
are shown in Table~\ref{table:pulse_phase}. 
Although the statistics of our data does not allow us to  
confirm the positive correlation between $E_{\rm cyc}$ and pulse amplitude
reported in  \citet{2011A&A...532A.126K} (although, some
support to the correlation is leaned by the \emph{RXTE} data), 
the measured slopes are in agreement, within uncertainties, with those
reported by \citet{2011A&A...532A.126K} and repeated in the bottom
section of Table~\ref{table:pulse_phase}

\section{Discussion}
\label{sec:discussion}

We have analyzed the phase-averaged, phase-resolved and pulse-to-pulse spectra 
of A\,0535+26 as a function of the source luminosity. 
In all types of the analysis, the continuum gets harder with increasing flux.
In some pulse-phase intervals, the data show indications of a positive
$E_\mathrm{cyc}$-flux correlation, although with marginal significance ($\sim$3\,$\sigma$).
The pulse-averaged centroid energy of the cyclotron line measured over longer
time scales (days) shows some irregular variability but no clear dependence
on flux, which is consistent with previous observations of this source.
In the following, we discuss our results in more detail.

\subsection{Long-term spectral variations with luminosity}
\label{sec:disc:longterm}

As mentioned in the introduction, a relation between the cyclotron line 
centroid energy and luminosity was reported for a number of accreting pulsars.  
Since the work of \citet{1976MNRAS.175..395B}, it is known that the 
geometry of the emitting structure (column), on/above the polar caps of an 
accreting neutron star, must be different for the luminosities above and below
a certain critical value $L_{\rm c}$ which depends on the pulsar's properties.
In the following, accreting pulsars with luminosities below
  $L_{\rm c}$ are referred to as \emph{sub-critical} sources while those
 with luminosities exceeding $L_{\rm c}$ -- as \emph{super-critical}.
According to this picture, V\,0332+53 and 4U\,0115+63, which show a 
negative $E_\mathrm{cyc}-L$ correlation and a softening of the spectrum with 
flux, belong to so-called \emph{super-critical} sources, whose luminosities
are above $L_{\rm c}$. In such sources,
infalling matter is most probably decelerated in a radiative shock,
whose height is believed to increase with $L$, i.e., drift
towards an area with a lower $B$-field strength. Her X-1 and  GX\,304-1,
where a positive $E_\mathrm{cyc}-L$ correlation and a hardening of the 
spectrum with flux are observed,
are the examples of \emph{sub-critical} sources ($L<L_{\rm c}$), where
infalling matter is stopped by the Coulomb drag and
collective plasma effects, rather than in a radiative shock.
In such sources, the height of the emitting region
decreases with increasing luminosity, owing to a corresponding
increase in ram pressure of the infalling material 
(\citealt{2007A&A...465L..25S} and \citealt{2012A&A...544A.123B}).

The newest theoretical calculations by \citet{2012A&A...544A.123B} predict
for A\,0535+26, a value of the critical luminosity 
$L_{\rm c}\sim 6.78\times10^{37}$\,erg/s, which is $\sim$5--6 times larger
than the peak flux of the outburst analyzed here (see the upper X-axis
of Fig.\,\ref{figure:2}). Thus, A\,0535+26 must have been accreting
in a sub-critical regime during the outburst. Though, the key indication
of this regime, namely the positive correlation between the CRSF energy and
flux, is not firmly established in this source. Its X-ray continuum shows,
however, a hardening with increasing flux, which is also observed
in the other two sub-critical sources, Her X-1 and GX\,304-1 (the
super-critical sources V\,0332+53 and 4U\,0115+63 show an opposite 
behavior of the spectral continuum, see above).
If the accretion regime in A\,0535+26 is similar to one in Her X-1 and
GX\,304-1, the expected positive $E_{\rm cyc}-L$ correlation must be
smeared due to a special geometry of the emitting structure and/or a special 
orientation of the system with respect to the observer's line of sight.
Alternatively, A\,0535+26 might represent yet another, ``third'' accretion
regime which yields very little, if any, variations in the height of
the emitting region above the star surface (i.e., in the measured
cyclotron line energy) in the 
observed range of luminosities. Indeed, while the luminosities
of Her X-1 and GX\,304-1 are below, but relatively close to the corresponding
values of $L_{\rm c}$ (see Fig.\,2 in \citealt{2012A&A...544A.123B}),
A\,0535+26 is far below $L_{\rm c}$. Thus, the configuration of the
accretion column in this source might differ substantially from those in
the other two sub-critical pulsars. These considerations seem to be
supported by the latest observations of a \emph{giant} outburst of
A\,0535+26, analyzed by \citet{2012AIPC.1427..300C}, when the source luminosity
approached $L_{\rm c}$ (i.e., a luminosity range of Her X-1 and GX\,304-1). 
In this preliminary analysis, a few measurements of $E_{\rm cyc}$, corresponding
to the highest luminosities, indeed show an indication of an increase towards
higher fluxes (Fig.\,1 of \citealt{2012AIPC.1427..300C}). Further observations
are, however, necessary to confirm this effect.

\subsection{Pulse-phase dependence of the spectrum}

It is 
necessary to reconstruct the geometry of the rotating neutron star 
and its emitting regions to understand the results of the phase-resolved spectral analysis. Such a reconstruction can be made based
on the observed pulse profiles of a pulsar using, e.g., the
decomposition method by \citet{1995ApJ...450..763K}.
This technique has recently been applied to A\,0535+26 by 
\citet{2011A&A...526A.131C}. As a result, the pulse profiles of the source
were reproduced in a model, including a hollow column, a thermal halo and 
taking into account relativistic light deflection in the strong gravitational
field of the neutron star. Derived from the results of \citet{2011A&A...526A.131C},
the observed pulse profile can be produced by
one magnetic pole passing through the observer's line of sight, 
whereas the second pole is never seen directly from the top but only 
through the gravitationally bent radiation from the lateral emission 
of the accretion column and its thermal scattering halo.

Since we observed similar pulse profiles as those used by 
\citet{2011A&A...526A.131C}, we adopt the reconstructed geometry
for our data.
In our analysis, we observe a slightly harder spectrum for the main pulse 
with respect to the secondary pulse. 
Magnetized plasma gets optically thin for photons (E$<E_{\mathrm{cyc}}$) 
which propagate parallel to the magnetic field lines. 
Thus, the main peak is probably formed by harder photons escaping
the accretion column from top, along the field lines.
About half a phase later, we probably see the lateral emission of both the 
accretion column and their thermal halos. 

In our phase-resolved analysis, we observe variations of the cyclotron line 
energy with pulse phase in range $\sim$42--50\,keV, which is probably caused
by the changing viewing conditions of the scattering region and
a dependence of the magnetic scattering cross sections on the 
angle between the magnetic field and the line of sight 
\citep{1991ApJ...374..687H}. 
Variations over pulse phase are also seen in the slope of the linear fits to 
the $E_{\rm cyc}$ and $\Gamma$ values as a function of flux 
(Fig.\,\ref{figure:6}). 
While a significant decrease of $\Gamma$ with flux is present at most
pulse phases, a considerable (positive) $E_{\rm cyc}$-flux correlation
is only seen during the secondary peak (Figs.\,\ref{figure:6} and
\ref{figure:phase_01_05_06_10_fit}). According to the geometrical
picture described above, the emerging radiation
is emitted roughly perpendicular to the axis of the accretion column(s), that
is, perpendicular to the field lines, in the secondary peak. As discussed in e.g. \citet{2011A&A...532A..76F},
at such angle, the resonant scattering features are expected
to be deeper and narrower. Therefore, in the secondary peak, the
viewing conditions of the line-forming region might be favourable for
measuring $E_{\rm cyc}$ variations with flux. In the
pulse-averaged spectral analysis, the dependence is smeared out.

\subsection{Pulse-to-pulse variability}

As we noted in Sect.\,\ref{sec:pulse_to_pulse_analysis}, the results
of our pulse-to-pulse analysis, namely the spectral variations with
the amplitude of individual pulses, are in agreement with those
reported by \citet{2011A&A...532A.126K} on the basis of the
observations of the source's outburst in 2005 with better photon
statistics. With our data, we have reproduced the hardening of the
power-law with pulse amplitude. An indication of a positive
correlation of the cyclotron line energy with flux can only be seen
in the \emph{RXTE} data (Fig.\,\ref{figure:12}). If the positive
correlation between the line energy and flux on the pulse-to-pulse time
scale in A\,0535+26 is real, the question remains: why the correlation
is not observed on the pulse-averaged spectra accumulated over longer
time scales? A qualitative explanation of this discrepancy might be the following.
Strong pulse-to-pulse variability of the source's
flux leads to the relatively broad distribution of the amplitudes of
individual pulses -- more than a factor of two, 
as indicated in
Fig.\,\ref{figure:0912_pulse_height_distribution}. Thus,
pulse-averaged spectra accumulated over a longer time include
contributions by pulses of very different heights. The pulse-to-pulse
variability of the source's spectrum must lead to a substantial
smearing effect, especially in case of a line-like feature. If
A\,0535+26 indeed accrets in a regime with very little (if any)
variation in the height of the emitting region with flux, as discussed
in Sect.\,\ref{sec:disc:longterm}, the reduced sensitivity of the
pulse-averaged spectroscopy to the $E_{\rm cyc}$--flux correlation might
indeed lead to the observed discrepancy between the pulse-averaged and
pulse-to-pulse results.

\section{Summary and conclusions}
\label{sec:summary_conslusions}

In this work, we present the spectral analysis of the observational
data taken with \emph{RXTE} and \emph{INTEGRAL}
on the Be/X-ray binary system A\,0535+26, during its outburst in April 2010. 
We studied the source's broad-band X-ray spectrum as a function of
luminosity on the pulse-averaged, pulse-phase resolved, and
pulse-to-pulse basis. In all cases, a hardening of the X-ray
continuum with flux is observed. This behavior is similar to that of
other sub-critical sources, Her X-1 and GX\,304-1. An indication of
a positive correlation between the cyclotron line energy and flux (a
characteristic feature of a sub-critical source) is only found in some
pulse-phase resolved and pulse-to-pulse spectra (Fig. \ref{figure:6} and Table~\ref{table:pulse_phase}). 

Based on the spectrum-flux dependence of A\,0535+26, which is similar
to that of other sub-critical sources, and on the source's luminosity,
which is well below the critical luminosity $L_{\rm c}$, we conclude
that the pulsar operates in a sub-critical accretion regime, where
the radiative shock, if formed, is not capable of decelerating the
accreted plasma to rest. Instead, the in-falling matter
is most probably stopped by the Coulomb drag and collective plasma effects. 
A similar behavior was reported in earlier works on this source, see e.g. \citet{2007A&A...465L..21C,2008A&A...480L..17C,2011arXiv1107.3417C}.

The lack of a firm detection of a $E_{\rm cyc}$--flux correlation,
which is expected in this accretion regime, we tentatively attribute to a substantially lower ratio of
the source luminosity to the critical one, compared to other
sub-critical sources. At such a low luminosity level, the amplitude of variation in
the height of the line-forming region with luminosity (mass accretion rate) might be
substantially narrower than that in Her X-1 and GX\,304-1.

\begin{acknowledgements} This research is based on observations with \emph{INTEGRAL}, an \emph{ESA} project with instruments and science data centre funded by \emph{ESA} member states (especially the PI countries: Denmark, France, Germany, Italy, Switzerland, Spain), Poland and with the participation of Russia and the USA. For this analysis we also used \emph{RXTE} observations provided by the RXTE Science Operations Center and \emph{Swift/BAT} transient monitor results provided by the Swift/BAT team. This work has been partially funded by the DLR, grant 50 OR 1008, and by the Carl-Zeiss-Stiftung. IC acknowledges financial support from the French Space Agency CNES through CNRS. We thank the anonymous referee for very valuable comments.\end{acknowledgements}

\bibliographystyle{aa}
\bibliography{articles}

\begin{thebibliography}{54}
\expandafter\ifx\csname natexlab\endcsname\relax\def\natexlab#1{#1}\fi

\bibitem[{{Acciari} {et~al.}(2011){Acciari}, {Aliu}, {Araya}, {Arlen}, {Aune},
  {Beilicke}, {Benbow}, {Bradbury}, {Buckley}, {Bugaev}, {Byrum}, {Cannon},
  {Cesarini}, {Ciupik}, {Collins-Hughes}, {Cui}, {Dickherber}, {Duke},
  {Falcone}, {Finley}, {Fortson}, {Furniss}, {Galante}, {Gall}, {Godambe},
  {Griffin}, {Guenette}, {Gyuk}, {Hanna}, {Holder}, {Hughes}, {Hui},
  {Humensky}, {Imran}, {Kaaret}, {Kertzman}, {Krawczynski}, {Krennrich},
  {Madhavan}, {Maier}, {Majumdar}, {McArthur}, {Moriarty}, {Ong}, {Otte},
  {Pandel}, {Park}, {Perkins}, {Pohl}, {Prokoph}, {Quinn}, {Ragan}, {Reyes},
  {Reynolds}, {Roache}, {Rose}, {Saxon}, {Sembroski}, {{\c S}ent{\"u}rk},
  {Smith}, {Te{\v s}i{\'c}}, {Theiling}, {Thibadeau}, {Varlotta}, {Vincent},
  {Vivier}, {Wakely}, {Ward}, {Weekes}, {Weinstein}, {Weisgarber}, {Weng},
  {Williams}, {Wood}, \& {Zitzer}}]{2011ApJ...733...96A}
{Acciari}, V.~A., {Aliu}, E., {Araya}, M., {et~al.} 2011, \apj, 733, 96

\bibitem[{{Araya} \& {Harding}(1996)}]{1996A&AS..120C.183A}
{Araya}, R.~A. \& {Harding}, A.~K. 1996, \aaps, 120, C183

\bibitem[{{Basko} \& {Sunyaev}(1976)}]{1976MNRAS.175..395B}
{Basko}, M.~M. \& {Sunyaev}, R.~A. 1976, \mnras, 175, 395

\bibitem[{{Becker} {et~al.}(2012){Becker}, {Klochkov}, {Sch{\"o}nherr},
  {Nishimura}, {Ferrigno}, {Caballero}, {Kretschmar}, {Wolff}, {Wilms}, \&
  {Staubert}}]{2012A&A...544A.123B}
{Becker}, P.~A., {Klochkov}, D., {Sch{\"o}nherr}, G., {et~al.} 2012, \aap, 544,
  A123

\bibitem[{{Bradt} {et~al.}(1993){Bradt}, {Rothschild}, \&
  {Swank}}]{1993A&AS...97..355B}
{Bradt}, H.~V., {Rothschild}, R.~E., \& {Swank}, J.~H. 1993, \aaps, 97, 355

\bibitem[{{Caballero}(2009)}]{2009PhDT........12C}
{Caballero}, I. 2009, PhD thesis, IAAT University of Tuebingen

\bibitem[{{Caballero} {et~al.}(2011{\natexlab{a}}){Caballero}, {Ferrigno},
  {Klochkov}, {Santangelo}, {Staubert}, {Kretschmar}, {Pottschmidt},
  {Kreykenbohm}, {Wilms}, {Postnov}, {Schoenherr}, {Rothschild}, {Suchy},
  {Finger}, \& {Camero-Arranz}}]{2011ATel.3204....1C}
{Caballero}, I., {Ferrigno}, C., {Klochkov}, D., {et~al.} 2011{\natexlab{a}},
  The Astronomer's Telegram, 3204, 1

\bibitem[{{Caballero} {et~al.}(2011{\natexlab{b}}){Caballero}, {Kraus},
  {Santangelo}, {Sasaki}, \& {Kretschmar}}]{2011A&A...526A.131C}
{Caballero}, I., {Kraus}, U., {Santangelo}, A., {Sasaki}, M., \& {Kretschmar},
  P. 2011{\natexlab{b}}, \aap, 526, A131

\bibitem[{{Caballero} {et~al.}(2009){Caballero}, {Kretschmar}, {Pottschmidt},
  {Wilms}, {Kreykenbohm}, {Suchy}, {Rothschild}, {Ferrigno}, {Santangelo},
  {Klochkov}, \& {Staubert}}]{2009ATel.2337....1C}
{Caballero}, I., {Kretschmar}, P., {Pottschmidt}, K., {et~al.} 2009, The
  Astronomer's Telegram, 2337, 1

\bibitem[{{Caballero} {et~al.}(2007){Caballero}, {Kretschmar}, {Santangelo},
  {Staubert}, {Klochkov}, {Camero}, {Ferrigno}, {Finger}, {Kreykenbohm},
  {McBride}, {Pottschmidt}, {Rothschild}, {Sch{\"o}nherr}, {Segreto}, {Suchy},
  {Wilms}, \& {Wilson}}]{2007A&A...465L..21C}
{Caballero}, I., {Kretschmar}, P., {Santangelo}, A., {et~al.} 2007, \aap, 465,
  L21

\bibitem[{{Caballero} {et~al.}(2012){Caballero}, {M{\"u}ller}, {Bordas},
  {Ferrigno}, {K{\"u}hnel}, {Pottschmidt}, {Kretschmar}, {Wilms},
  {Kreykenbohm}, {Klochkov}, {Santangelo}, {Staubert}, {Suchy}, {Rothschild},
  {Camero-Arranz}, \& {Finger}}]{2012AIPC.1427..300C}
{Caballero}, I., {M{\"u}ller}, S., {Bordas}, P., {et~al.} 2012, in American
  Institute of Physics Conference Series, Vol. 1427, American Institute of
  Physics Conference Series, ed. R.~{Petre}, K.~{Mitsuda}, \& L.~{Angelini},
  300--301

\bibitem[{{Caballero} {et~al.}(2013){Caballero}, {Pottschmidt}, {Marcu},
  {Barragan}, {Ferrigno}, {Klochkov}, {Zurita Heras}, {Suchy}, {Wilms},
  {Kretschmar}, {Santangelo}, {Kreykenbohm}, {F{\"u}rst}, {Rothschild},
  {Staubert}, {Finger}, {Camero-Arranz}, {Makishima}, {Enoto}, {Iwakiri}, \&
  {Terada}}]{2013ApJ...764...L23}
{Caballero}, I., {Pottschmidt}, K., {Marcu}, D.~M., {et~al.} 2013, \apj, 764,
  L23

\bibitem[{{Caballero} {et~al.}(2011{\natexlab{c}}){Caballero}, {Pottschmidt},
  {Santangelo}, {Barragan}, {Klochkov}, {Ferrigno}, {Rodriguez}, {Kretschmar},
  {Suchy}, {Marcu}, {Mueller}, {Wilms}, {Kreykenbohm}, {Rothschild},
  {Staubert}, {Finger}, {Camero-Arranz}, {Makishima}, {Mihara}, {Nakajima},
  {Enoto}, {Iwakiri}, \& {Terada}}]{2011arXiv1107.3417C}
{Caballero}, I., {Pottschmidt}, K., {Santangelo}, A., {et~al.}
  2011{\natexlab{c}}, ArXiv e-prints

\bibitem[{{Caballero} {et~al.}(2008){Caballero}, {Santangelo}, {Kretschmar},
  {Staubert}, {Postnov}, {Klochkov}, {Camero-Arranz}, {Finger}, {Kreykenbohm},
  {Pottschmidt}, {Rothschild}, {Suchy}, {Wilms}, \&
  {Wilson}}]{2008A&A...480L..17C}
{Caballero}, I., {Santangelo}, A., {Kretschmar}, P., {et~al.} 2008, \aap, 480,
  L17

\bibitem[{{Caballero} {et~al.}(2010){Caballero}, {Santangelo}, {Pottschmidt},
  {Klochkov}, {Rodriguez}, {Wilms}, {Kreykenbohm}, {Kretschmar}, {Ferrigno},
  {Rothschild}, \& {Suchy}}]{2010ATel.2541....1C}
{Caballero}, I., {Santangelo}, A., {Pottschmidt}, K., {et~al.} 2010, The
  Astronomer's Telegram, 2541, 1

\bibitem[{{Camero-Arranz} {et~al.}(2011){Camero-Arranz}, {Finger},
  {Wilson-Hodge}, {Jenke}, {Steele}, {Gutierrez-Soto}, \&
  {Coe}}]{2011ATel.3166....1C}
{Camero-Arranz}, A., {Finger}, M.~H., {Wilson-Hodge}, C., {et~al.} 2011, The
  Astronomer's Telegram, 3166, 1

\bibitem[{{Ferrigno} {et~al.}(2011){Ferrigno}, {Falanga}, {Bozzo}, {Becker},
  {Klochkov}, \& {Santangelo}}]{2011A&A...532A..76F}
{Ferrigno}, C., {Falanga}, M., {Bozzo}, E., {et~al.} 2011, \aap, 532, A76

\bibitem[{{Finger} {et~al.}(2006){Finger}, {Camero-Arranz}, {Kretschmar},
  {Wilson}, \& {Patel}}]{2006HEAD....9.0759F}
{Finger}, M.~H., {Camero-Arranz}, A., {Kretschmar}, P., {Wilson}, C., \&
  {Patel}, S. 2006, in Bulletin of the American Astronomical Society, Vol.~38,
  AAS/High Energy Astrophysics Division \#9, 359

\bibitem[{{Finger} {et~al.}(1996){Finger}, {Wilson}, \&
  {Harmon}}]{1996ApJ...459..288F}
{Finger}, M.~H., {Wilson}, R.~B., \& {Harmon}, B.~A. 1996, \apj, 459, 288

\bibitem[{{Giangrande} {et~al.}(1980){Giangrande}, {Giovannelli}, {Bartolini},
  {Guarnieri}, \& {Piccioni}}]{1980A&AS...40..289G}
{Giangrande}, A., {Giovannelli}, F., {Bartolini}, C., {Guarnieri}, A., \&
  {Piccioni}, A. 1980, \aaps, 40, 289

\bibitem[{{Grove} {et~al.}(1995){Grove}, {Strickman}, {Johnson}, {Kurfess},
  {Kinzer}, {Starr}, {Jung}, {Kendziorra}, {Kretschmar}, {Maisack}, \&
  {Staubert}}]{1995ApJ...438L..25G}
{Grove}, J.~E., {Strickman}, M.~S., {Johnson}, W.~N., {et~al.} 1995, \apjl,
  438, L25

\bibitem[{{Harding} \& {Daugherty}(1991)}]{1991ApJ...374..687H}
{Harding}, A.~K. \& {Daugherty}, J.~K. 1991, \apj, 374, 687

\bibitem[{{Isenberg} {et~al.}(1998){Isenberg}, {Lamb}, \&
  {Wang}}]{1998ApJ...493..154I}
{Isenberg}, M., {Lamb}, D.~Q., \& {Wang}, J.~C.~L. 1998, \apj, 493, 154

\bibitem[{{Jahoda} {et~al.}(2006){Jahoda}, {Markwardt}, {Radeva}, {Rots},
  {Stark}, {Swank}, {Strohmayer}, \& {Zhang}}]{2006ApJS..163..401J}
{Jahoda}, K., {Markwardt}, C.~B., {Radeva}, Y., {et~al.} 2006, \apjs, 163, 401

\bibitem[{{Kendziorra} {et~al.}(1994){Kendziorra}, {Kretschmar}, {Pan}, {Kunz},
  {Maisack}, {Staubert}, {Pietsch}, {Truemper}, {Efremov}, \&
  {Sunyaev}}]{1994A&A...291L..31K}
{Kendziorra}, E., {Kretschmar}, P., {Pan}, H.~C., {et~al.} 1994, \aap, 291, L31

\bibitem[{{Klochkov} {et~al.}(2012){Klochkov}, {Doroshenko}, {Santangelo},
  {Staubert}, {Ferrigno}, {Kretschmar}, {Caballero}, {Wilms}, {Kreykenbohm},
  {Pottschmidt}, {Rothschild}, {Wilson-Hodge}, \&
  {P{\"u}hlhofer}}]{2012A&A...542L..28K}
{Klochkov}, D., {Doroshenko}, V., {Santangelo}, A., {et~al.} 2012, \aap, 542,
  L28

\bibitem[{{Klochkov} {et~al.}(2011){Klochkov}, {Staubert}, {Santangelo},
  {Rothschild}, \& {Ferrigno}}]{2011A&A...532A.126K}
{Klochkov}, D., {Staubert}, R., {Santangelo}, A., {Rothschild}, R.~E., \&
  {Ferrigno}, C. 2011, \aap, 532, A126

\bibitem[{{Kraus} {et~al.}(1995){Kraus}, {Nollert}, {Ruder}, \&
  {Riffert}}]{1995ApJ...450..763K}
{Kraus}, U., {Nollert}, H.-P., {Ruder}, H., \& {Riffert}, H. 1995, \apj, 450,
  763

\bibitem[{{Krimm} {et~al.}(2009){Krimm}, {Barthelmy}, {Baumgartner},
  {Cummings}, {Fenimore}, {Gehrels}, {Markwardt}, {Palmer}, {Sakamoto},
  {Skinner}, {Stamatikos}, {Tueller}, \& {Ukwatta}}]{2009ATel.2336....1K}
{Krimm}, H.~A., {Barthelmy}, S.~D., {Baumgartner}, W., {et~al.} 2009, The
  Astronomer's Telegram, 2336, 1

\bibitem[{{Lebrun} {et~al.}(2003){Lebrun}, {Leray}, {Lavocat}, {Cr{\'e}tolle},
  {Arqu{\`e}s}, {Blondel}, {Bonnin}, {Bou{\`e}re}, {Cara}, {Chaleil}, {Daly},
  {Desages}, {Dzitko}, {Horeau}, {Laurent}, {Limousin}, {Mathy}, {Mauguen},
  {Meignier}, {Molini{\'e}}, {Poindron}, {Rouger}, {Sauvageon}, \&
  {Tourrette}}]{2003A&A...411L.141L}
{Lebrun}, F., {Leray}, J.~P., {Lavocat}, P., {et~al.} 2003, \aap, 411, L141

\bibitem[{{Li} {et~al.}(1979){Li}, {Clark}, {Jernigan}, \&
  {Rappaport}}]{1979ApJ...228..893L}
{Li}, F., {Clark}, G.~W., {Jernigan}, J.~G., \& {Rappaport}, S. 1979, \apj,
  228, 893

\bibitem[{{Lund} {et~al.}(2003){Lund}, {Budtz-J{\o}rgensen}, {Westergaard},
  {Brandt}, {Rasmussen}, {Hornstrup}, {Oxborrow}, {Chenevez}, {Jensen},
  {Laursen}, {Andersen}, {Mogensen}, {Rasmussen}, {Om{\o}}, {Pedersen},
  {Polny}, {Andersson}, {Andersson}, {K{\"a}m{\"a}r{\"a}inen}, {Vilhu},
  {Huovelin}, {Maisala}, {Morawski}, {Juchnikowski}, {Costa}, {Feroci},
  {Rubini}, {Rapisarda}, {Morelli}, {Carassiti}, {Frontera}, {Pelliciari},
  {Loffredo}, {Mart{\'{\i}}nez N{\'u}{\~n}ez}, {Reglero}, {Velasco}, {Larsson},
  {Svensson}, {Zdziarski}, {Castro-Tirado}, {Attina}, {Goria}, {Giulianelli},
  {Cordero}, {Rezazad}, {Schmidt}, {Carli}, {Gomez}, {Jensen}, {Sarri},
  {Tiemon}, {Orr}, {Much}, {Kretschmar}, \& {Schnopper}}]{2003A&A...411L.231L}
{Lund}, N., {Budtz-J{\o}rgensen}, C., {Westergaard}, N.~J., {et~al.} 2003,
  \aap, 411, L231

\bibitem[{{Makino} {et~al.}(1989){Makino}, {Cook}, {Grunsfeld}, {Heindl},
  {Palmer}, {Prince}, {Schindler}, {Stone}, \& {Sunyaev}}]{1989IAUC.4769....1M}
{Makino}, F., {Cook}, W., {Grunsfeld}, J., {et~al.} 1989, \iaucirc, 4769, 1

\bibitem[{{Motch} {et~al.}(1991){Motch}, {Stella}, {Janot-Pacheco}, \&
  {Mouchet}}]{1991ApJ...369..490M}
{Motch}, C., {Stella}, L., {Janot-Pacheco}, E., \& {Mouchet}, M. 1991, \apj,
  369, 490

\bibitem[{{Mowlavi} {et~al.}(2006){Mowlavi}, {Kreykenbohm}, {Shaw},
  {Pottschmidt}, {Wilms}, {Rodriguez}, {Produit}, {Soldi}, {Larsson}, \&
  {Dubath}}]{2006A&A...451..187M}
{Mowlavi}, N., {Kreykenbohm}, I., {Shaw}, S.~E., {et~al.} 2006, \aap, 451, 187

\bibitem[{{Nagase} {et~al.}(1982){Nagase}, {Hayakawa}, {Kunieda}, {Makino},
  {Masai}, {Tawara}, {Inoue}, {Kawai}, {Koyama}, {Makishima}, {Matsuoka},
  {Murakami}, {Oda}, {Ogawara}, {Ohashi}, {Shibazaki}, {Tanaka}, {Miyamoto},
  {Tsunemi}, {Yamashita}, \& {Kondo}}]{1982ApJ...263..814N}
{Nagase}, F., {Hayakawa}, S., {Kunieda}, H., {et~al.} 1982, \apj, 263, 814

\bibitem[{{Pottschmidt} {et~al.}(2006){Pottschmidt}, {Rothschild}, {Gasaway},
  {Suchy}, \& {Coburn}}]{2006HEAD....9.1821P}
{Pottschmidt}, K., {Rothschild}, R.~E., {Gasaway}, T., {Suchy}, S., \&
  {Coburn}, W. 2006, in Bulletin of the American Astronomical Society, Vol.~38,
  AAS/High Energy Astrophysics Division \#9, 384--+

\bibitem[{{Ricketts} {et~al.}(1975){Ricketts}, {Turner}, {Page}, \&
  {Pounds}}]{1975Natur.256..631R}
{Ricketts}, M.~J., {Turner}, M.~J.~L., {Page}, C.~G., \& {Pounds}, K.~A. 1975,
  \nat, 256, 631

\bibitem[{{Rosenberg} {et~al.}(1975){Rosenberg}, {Eyles}, {Skinner}, \&
  {Willmore}}]{1975Natur.256..628R}
{Rosenberg}, F.~D., {Eyles}, C.~J., {Skinner}, G.~K., \& {Willmore}, A.~P.
  1975, \nat, 256, 628

\bibitem[{{Rothschild} {et~al.}(1998){Rothschild}, {Blanco}, {Gruber},
  {Heindl}, {MacDonald}, {Marsden}, {Pelling}, {Wayne}, \&
  {Hink}}]{1998ApJ...496..538R}
{Rothschild}, R.~E., {Blanco}, P.~R., {Gruber}, D.~E., {et~al.} 1998, \apj,
  496, 538

\bibitem[{{Rothschild} {et~al.}(2011){Rothschild}, {Markowitz}, {Rivers},
  {Suchy}, {Pottschmidt}, {Kadler}, {M{\"u}ller}, \&
  {Wilms}}]{2011ApJ...733...23R}
{Rothschild}, R.~E., {Markowitz}, A., {Rivers}, E., {et~al.} 2011, \apj, 733,
  23

\bibitem[{{Rothschild} {et~al.}(2006){Rothschild}, {Wilms}, {Tomsick},
  {Staubert}, {Benlloch}, {Collmar}, {Madejski}, {Deluit}, \&
  {Khandrika}}]{2006ApJ...641..801R}
{Rothschild}, R.~E., {Wilms}, J., {Tomsick}, J., {et~al.} 2006, \apj, 641, 801

\bibitem[{{Sembay} {et~al.}(1990){Sembay}, {Schwartz}, {Orwig}, {Dennis}, \&
  {Davies}}]{1990ApJ...351..675S}
{Sembay}, S., {Schwartz}, R.~A., {Orwig}, L.~E., {Dennis}, B.~R., \& {Davies},
  S.~R. 1990, \apj, 351, 675

\bibitem[{{Staubert} {et~al.}(2007){Staubert}, {Shakura}, {Postnov}, {Wilms},
  {Rothschild}, {Coburn}, {Rodina}, \& {Klochkov}}]{2007A&A...465L..25S}
{Staubert}, R., {Shakura}, N.~I., {Postnov}, K., {et~al.} 2007, \aap, 465, L25

\bibitem[{{Steele} {et~al.}(1998){Steele}, {Negueruela}, {Coe}, \&
  {Roche}}]{1998MNRAS.297L...5S}
{Steele}, I.~A., {Negueruela}, I., {Coe}, M.~J., \& {Roche}, P. 1998, \mnras,
  297, L5+

\bibitem[{{Stella} {et~al.}(1986){Stella}, {White}, \&
  {Rosner}}]{1986ApJ...308..669S}
{Stella}, L., {White}, N.~E., \& {Rosner}, R. 1986, \apj, 308, 669

\bibitem[{{Tchernin} {et~al.}(2011){Tchernin}, {Ferrigno}, \&
  {Bozzo}}]{2011ATel.3173....1T}
{Tchernin}, C., {Ferrigno}, C., \& {Bozzo}, E. 2011, The Astronomer's Telegram,
  3173, 1

\bibitem[{{Truemper} {et~al.}(1978){Truemper}, {Pietsch}, {Reppin}, {Voges},
  {Staubert}, \& {Kendziorra}}]{1978ApJ...219L.105T}
{Truemper}, J., {Pietsch}, W., {Reppin}, C., {et~al.} 1978, \apjl, 219, L105

\bibitem[{{Tsygankov} {et~al.}(2006){Tsygankov}, {Lutovinov}, {Churazov}, \&
  {Sunyaev}}]{2006MNRAS.371...19T}
{Tsygankov}, S.~S., {Lutovinov}, A.~A., {Churazov}, E.~M., \& {Sunyaev}, R.~A.
  2006, \mnras, 371, 19

\bibitem[{{Tsygankov} {et~al.}(2007){Tsygankov}, {Lutovinov}, {Churazov}, \&
  {Sunyaev}}]{2007AstL...33..368T}
{Tsygankov}, S.~S., {Lutovinov}, A.~A., {Churazov}, E.~M., \& {Sunyaev}, R.~A.
  2007, Astronomy Letters, 33, 368

\bibitem[{{Tsygankov} {et~al.}(2010){Tsygankov}, {Lutovinov}, \&
  {Serber}}]{2010MNRAS.401.1628T}
{Tsygankov}, S.~S., {Lutovinov}, A.~A., \& {Serber}, A.~V. 2010, \mnras, 401,
  1628

\bibitem[{{Tueller} {et~al.}(2005){Tueller}, {Ajello}, {Barthelmy}, {Krimm},
  {Makwardt}, \& {Skinner}}]{2005ATel..504....1T}
{Tueller}, J., {Ajello}, M., {Barthelmy}, S., {et~al.} 2005, The Astronomer's
  Telegram, 504, 1

\bibitem[{{Ubertini} {et~al.}(2003){Ubertini}, {Lebrun}, {Di Cocco}, {Bazzano},
  {Bird}, {Broenstad}, {Goldwurm}, {La Rosa}, {Labanti}, {Laurent}, {Mirabel},
  {Quadrini}, {Ramsey}, {Reglero}, {Sabau}, {Sacco}, {Staubert}, {Vigroux},
  {Weisskopf}, \& {Zdziarski}}]{2003A&A...411L.131U}
{Ubertini}, P., {Lebrun}, F., {Di Cocco}, G., {et~al.} 2003, \aap, 411, L131

\bibitem[{{Winkler} {et~al.}(2003){Winkler}, {Courvoisier}, {Di Cocco},
  {Gehrels}, {Gim{\'e}nez}, {Grebenev}, {Hermsen}, {Mas-Hesse}, {Lebrun},
  {Lund}, {Palumbo}, {Paul}, {Roques}, {Schnopper}, {Sch{\"o}nfelder},
  {Sunyaev}, {Teegarden}, {Ubertini}, {Vedrenne}, \&
  {Dean}}]{2003A&A...411L...1W}
{Winkler}, C., {Courvoisier}, T.~J.-L., {Di Cocco}, G., {et~al.} 2003, \aap,
  411, L1

\end{thebibliography}
\end{document}